\documentclass[aps,prd,preprintnumbers,amsmath,amssymb,nofootinbib,superscriptaddress,floatfix]{revtex4-2}
\usepackage[latin1]{inputenc}
\usepackage{slashed,xspace,nicefrac}
\usepackage{amsmath,amssymb,amsfonts}
\usepackage{textcomp}
\usepackage{natbib}
\usepackage{cancel}
\usepackage{enumitem}
\usepackage{array,multirow}
\usepackage{threeparttable}
\usepackage{appendix}
\usepackage[table,xcdraw]{xcolor}
\usepackage{color} 
\usepackage{tikz}
\usepackage{makecell}
\usepackage[normalem]{ulem}
\usepackage[unicode]{hyperref}
\newcommand{\be}{\begin{equation}}
\newcommand{\ee}{\end{equation}}
\newcommand{\bea}{\begin{eqnarray}}
\newcommand{\eea}{\end{eqnarray}}

\usepackage{graphicx,here}

\DeclareGraphicsExtensions{.pdf,.png,.jpg}
\graphicspath{{./}{figs/}{figures/}}
\begin{document}
\begin{flushleft} 
KCL-PH-TH/2026-{\bf 03}
\end{flushleft}

\title{The most general four-derivative Unitary String Effective Action with Torsion and Stringy-Running-Vacuum-Model Inflation: Old ideas from a modern perspective}

\author{Nick E. Mavromatos}
\affiliation{Physics Division, School of Applied Mathematical and Physical Sciences, National Technical University of Athens, 15780 Zografou Campus, Athens, Greece}
\affiliation{Theoretical Particle Physics and Cosmology Group, Department of Physics, King's College London, London, WC2R 2LS, UK}

\author{George Panagopoulos}
\email{gpanagopoulos@mail.ntua.gr}
\affiliation{Physics Division, School of Applied Mathematical and Physical Sciences, National Technical University of Athens, 15780 Zografou Campus,
Athens, Greece}

\begin{abstract}

The string-inspired running vacuum model (StRVM) of inflation is based on a
Chern--Simons (CS) gravity effective action in which the only
four-spacetime-derivative-order term is a gravitational anomalous CS--Pontryagin
density coupled to an axion. In this work, we revisit curvature-squared
string-inspired effective actions from the point of view of appropriate local
field redefinitions, leaving the perturbative string scattering matrices
invariant. We require simultaneously unitarity and torsion interpretation of
the field strength of the Kalb--Ramond antisymmetric tensor, features
characterizing the (3+1)-dimensional StRVM cosmology. Unlike the higher-dimensional case, the above features are possible in the context of
(3+1)-dimensional spacetimes, obtained after string compactification. We
demonstrate that the unitarity and torsion interpretation requirements lead to
a single type of extra four-derivative terms in the effective gravitational
action, not discussed in the previous literature on StRVM, which  is, however,
shown to be subleading by many orders of magnitude compared to the terms of
the StRVM framework. Hence, its presence has no practical implications for the
relevant inflationary (and, hence, postinflationary) physics of the StRVM. This
demonstrates the phenomenological completeness of the StRVM cosmological
scenario, which is thus fully embeddable in the UV-complete (quantum gravity-compatible) string theory framework.

\end{abstract}

\maketitle

\section{Introduction}

In a series of articles in the recent literature~\cite{bms0,bms,bms2,ms1,ms2,Mavromatos:2022xdo,Dorlis:2024yqw,Dorlis:2024uei}, a~model of inflation of the running vacuum model (RVM) type~\cite{SolaPeracaula:2026trz,SolaPeracaula:2025yco,SolaPeracaula:2023swx,Moreno-Pulido:2023ryo,SolaPeracaula:2022hpd,Sola:2015rra,Sola:2007sv,Moreno-Pulido:2022upl,Moreno-Pulido:2022phq} has been developed in the context of string theory, in~which inflation is produced by a condensate of the gravitational anomaly terms of the Chern--Simons (CS) type, characterizing the early Universe. This condensate is induced by primordial gravitational wave (GW) tensor perturbations. The~anomaly terms are remnants in (3+1)-spacetime dimensions of the Green--Schwarz counterterms in the string effective action, relevant to the anomaly cancellation mechanism~\cite{Green:1984sg}. In~the string-inspired RVM model (StRVM)~\cite{bms,ms1}, such terms are assumed present in the primordial Universe, whose dynamics is described only by fields in the massless gravitational multiplet of strings (which is also the ground state of the superstring)~\cite{Green_Schwarz_Witten_2012,Polchinski_1998}. Assuming 
a dynamical supergravity breaking mechanism in  the very early, pre-RVM-inflation-superstring Universe~\cite{bms,ms1,Basilakos:2015yoa,Gomez-Valent:2023hov} implies  
that the supersymmetry partners of the massless gravitational string multiplet become massive, with~masses near the Planck scale, thereby leaving behind as relevant degrees of freedom  the massless graviton, dilaton, and~spin-1 antisymmetric tensor (Kalb--Ramond (KR)) fields~\cite{Green_Schwarz_Witten_2012}, which drive  RVM inflation.
In the StRVM approach, the dilaton is self-consistently assumed to be stabilized to a constant value~\cite{bms2} that phenomenologically determines the string coupling that enters the (3+1)-dimensional gauge couplings of the low-energy effective field theory derived from the string. Thus, it plays no further role in the~cosmology.

In StRVM cosmology, the gravitational low-energy effective action is assumed to be of the lowest non-trivial order~\cite{Green_Schwarz_Witten_2012} in an  expansion in powers of the Regge slope $\alpha^\prime = M_s^{-2}$ (where $M_s$ is the string energy scale).
The Green--Schwarz anomaly counterterms~\cite{Green:1984sg}
enter through appropriate modifications of the field strength's three forms $\mathcal H_{\mu\nu\rho}$, $\mu, \nu, \rho =0, \dots, 3$   of the spin-1 KR antisymmetric tensor field in (3+1) dimensions after string compactification. These modifications imply a Bianchi identity for the field strength, which, when implemented in the path integral of the low-energy gravitational theory, after~path-integration with respect to the field strength, yields~\cite{Duncan:1992vz}, an effective gravitational action term containing a massless axion-like field (KR axion) $b(x)$ coupled to the CS gravitational anomaly terms (also called Hirzebruch  signature or gravitational Pontryagin density~\cite{EGUCHI1980213}).
This yields a CS-gravity~\cite{Jackiw:2003pm,Alexander:2009tp} effective field theory for the primordial StRVM~Universe.

The CS anomaly term is of quartic order in derivatives, being quadratic in curvature. When the dilaton is constant, a~graviton-ghost-free, and~thus unitary, Gauss--Bonnet (GB) term that is of quadratic-order in the curvature tensors 
can always be arranged to characterize the effective string action by appropriate graviton field redefinitions in perturbative string theory~\cite{Zwiebach:1985uq}. In~a (3+1)-dimensional  spacetime, the~GB combination 
is a total derivative~\cite{EGUCHI1980213} and thus plays no role in the physics of the StRVM, since its spacetime integral, and~hence its contribution to the effective action, vanishes upon~assuming appropriate boundary~conditions.

However, this may not be the end of the story. To~be formally consistent with the most general effective action, expanded up to and including ${\mathcal O}(\alpha^\prime)$ (four-spacetime-derivative) terms, one has to consider the corresponding $\mathcal O(\alpha^\prime)$ terms involving the field strength of the antisymmetric tensor field,
which are present in the string effective action~\cite{Gross:1986mw,Metsaev:1987zx,Bento:1986hx,Bern:1987bq}. 
It is in their presence that one, then, has to perform  the path integration over the $\mathcal H$-field strength so as to arrive at a form of the effective action involving a canonically normalized KR axion-like field. Classically, one may naively think that such a field can arise by identifying (in (3+1) dimensions) its spacetime derivative with the Hodge dual~\cite{EGUCHI1980213} of the three-form~$\mathcal H$:
\begin{align}\label{dual}
\partial_\sigma b \propto \eta_{\sigma\mu\nu\rho}\, \mathcal H^{\mu\nu\rho}\,,
\end{align}
with $\eta_{\mu\nu\rho\sigma}$ being the Levi--Civita density (see \eqref{levicivita}, below), 
and replacing it in the effective action~\cite{Duncan:1992vz}. This is not quite correct though, given that we want to establish the emergence of the axion at a full quantum level after~the $\mathcal H$-field path integration. In~principle, the~
fourth order in derivative effective action contains also quartic fields in the KR $H$-field strength, which would complicate obtaining analytic results after $\mathcal H$-path integration.  Moreover, as~we shall see in Section~\ref{sec:Haxion}, the~relation \eqref{dual} is modified once higher-order corrections are taken into account (cf. \eqref{dual3}).

It is the purpose of this work to first discuss the independent effective gravitational action terms in such an expansion, after~ appropriate \emph{local} field redefinitions, that leave the scattering (S-) matrix of perturbative string theory invariant. In~this respect, we shall re-examine (using a different basis of terms) the $\mathcal O(\alpha^\prime)$ terms of the bosonic part of the string effective action (heterotic~\cite{Green_Schwarz_Witten_2012,Polchinski_1998}, for~concreteness and phenomenological relevance), involving antisymmetric tensor and graviton fields~\cite{Gross:1986mw,Metsaev:1987zx,Bento:1986hx,Bern:1987bq}.
We shall impose target--space unitarity (i.e.,~Gauss--Bonnet-type curvature-squared terms) and torsion interpretation for the antisymmetric torsion field strength~\cite{Green_Schwarz_Witten_2012,Duncan:1992vz}. As~we shall see, unlike the higher $D$-dimensional ($D > (3+1)$) case of~\cite{Bento:1986hx,Bern:1987bq}, 
in the (3+1)-dimensional effective action (after string compactification), the imposition
of both requirements is compatible, which allows for a direct link with the StRVM framework~\cite{bms,ms1}. 
Then, we shall argue that there is a scheme in which the additional (compared to the StRVM action) four-derivative terms
are quadratic in the $\mathcal H$-field, contracted appropriately with Ricci and Ricci scalar curvature tensors, and~thus the $\mathcal H$-path integration remains Gaussian and can be performed analytically. In~this way, we arrive at a non-minimal coupling of $\propto \alpha^\prime \partial_\mu b \partial_\nu b $
terms,  with the Ricci tensor $R^{\mu\nu}$ as~the only type of extra term in the contorted unitary $\mathcal O(\alpha^\prime) $
effective action. In~cosmological settings, such terms
will be shown to be extremely suppressed compared to the terms kept in the StRVM analysis~\cite{bms0,bms,bms2,ms1,ms2,Mavromatos:2022xdo,Dorlis:2024yqw,Dorlis:2024uei}, and~hence, they do not affect the earlier conclusions on StRVM cosmology, which is, therefore, fully embeddable in the string~framework. 

The structure of this article is the following: in Section~\ref{sec:het}, we formulate the $\mathcal O(\alpha^\prime)$ (four-spacetime derivative) effective target spacetime action of the (heterotic, for~concreteness) string theory based on the massless bosonic string multiplet of graviton and antisymmetric tensor field strength $\mathcal H_{\mu\nu\rho}$, with~constant dilatons. In~Section~\ref{sec:redef}, we discuss  field redefinitions in the ten-dimensional target spacetime of strings that leave the perturbative target spacetime scattering (S-)matrix invariant under the assumptions of unitarity and (totally antisymmetric) torsion interpretation of $\mathcal H_{\mu\nu\rho}$ ($\mathcal H$-torsion). 
In Section~\ref{sec:Haxion}, we verify the well-known fact of the emergence of dynamical axion-like fields in the (3+1)-dimensional effective action after~string compactification, obtained by integrating out the $\mathcal H$-torsion in this more general context. The arising axions are, in this sense, dual to the $\mathcal H$-torsion in (3+1) dimensions, as~in the StRVM case, but~now, the (3+1)-dimensional string-inspired effective action contains extra terms, as~compared to the StRVM case, In Section~\ref{sec:strvminfl}, we discuss the effects of these extra terms and show that they do not affect the original conclusions on the inflationary era in the StRVM. 
Finally, Section~\ref{sec:concl} contains our conclusions. Some technical aspects of 
our work are presented in the four appendices. Specifically, Appendix \ref{sec:identities} discusses mathematical identities 
among generalized-curvature Riemann terms and~combinations thereof, with~totally antisymmetric torsion and~their decomposition into torsion-free and contorted parts, as~well as other useful identities among the torsion tensors used in our analysis. In~Appendix \ref{sec:beyondstrng}, we discuss, for~completeness, the~case where the field redefinitions are performed directly in a $D=4$ unitary effective field action with~a totally antisymmetric torsion that may be unrelated to a string theory one, and~we compare with the string theory case. 
In Appendix \ref{sec:parityandR2}, we make come comments on generalized contorted parity-odd curvature invariants, of~potential interest to gravitational anomalies, stressing their (trivial) behavior once the torsion is reduced to only its totally antisymmetric components, as~is the case of the string-inspired $\mathcal H$-torsion we are examining here.
Finally, Appendix \ref{sec:confgCS} includes 
technical details on the contributions of the new (as compared to the study in~\cite{bms,ms1,Dorlis:2024yqw}) terms of the ${\mathcal O}(\alpha^\prime)$ effective action to the gravitational CS anomaly~condensate.

\section{The Effective Action of Heterotic String~Theory}
\label{sec:het}

It is well known that the bosonic gravitational sector of a generic string theory consists of three massless fields~\cite{Green_Schwarz_Witten_2012}: a traceless, symmetric, dimensionless, spin-2 tensor field $g_{\mu \nu}$, uniquely identified with the graviton; a~dimensionless spin-0 (scalar) field $\Phi$, known as the dilaton; and~the dimensionless spin-1 antisymmetric (Kalb--Ramond) field $B_{\mu \nu}$ 
\footnote{For superstrings, the~above multiplet constitutes also the bosonic part of the respective ground state~\cite{Green_Schwarz_Witten_2012,Polchinski_1998}.}.

In the closed-string sector, there is a $U(1)$ gauge symmetry $B_{\mu \nu} \rightarrow B_{\mu \nu} + \partial_\mu \theta_\nu - \partial_\nu \theta_\mu$ and, as~such, the~low-energy string effective action depends solely on the field strength of the Kalb--Ramond field $B_{\mu \nu}$, as follows:
\begin{equation}
    H_{\mu \nu \rho} = \partial_{[\mu} B_{\nu \rho]},
\end{equation}
where the symbol $[\ldots]$ denotes complete antisymmetrization of the respective indices\footnote{Our normalization conventions for the antisymmetrization symbol $[\dots ]$ in indices are
$$\mathcal T_{[\mu_1 \dots \mu_n]} = \frac{1}{n!} \sum_{P \in \mathcal S_n} {\rm sign}(P)\, T_{\mu_{P(1)}\mu_{P(2)}\dots \mu_{P(n)}}\,,$$ 
where the symbol $P$ denotes permutations. Odd (even) permutations have sign $(P_{\rm odd (even)}) = -1 (+1)$.}.
The Kalb--Ramond field strength, which is a three-form, satisfies the Bianchi identity as follows:
\begin{equation}
    \partial_{[\mu} H_{\nu \rho \sigma]}=0,
\end{equation}
In other words,~its exterior derivative vanishes by~construction. In~heterotic strings, anomaly cancellation requirements dictate that Lorentz and gauge Chern--Simons terms be added to the field strength of the Kalb--Ramond field $B_{\mu \nu}$ (the so-called Green--Schwarz (GS) mechanism~\cite{Green:1984sg}), 
such that we use the language of differential forms~\cite{EGUCHI1980213} here, for~index economy reasons:
\begin{equation}\label{KRH}
    \boldsymbol{\mathcal{H}} = \boldsymbol{dB} + \frac{\alpha^\prime}{8 \kappa} (\Omega_{3L} - \Omega_{3Y}),
\end{equation}
where $\mathcal{H}_{\mu \nu \rho} = \kappa^{-1} H_{\mu \nu \rho}$ has dimensions of $[mass]^2$, 
$\kappa = M_{\rm Pl}^{-1}$ is the (3+1)-dimensional gravitational coupling, with~$M_{\rm Pl} = 2.435 \times 10^{18}~{\rm GeV}$ the reduced Planck mass, and 
$\alpha^\prime =M_s^{-2}$ is the Regge slope of the string, with~$M_s$ the string mass scale, which is not necessarily equal to the Planck mass scale. The~Lorentz and gauge Chern--Simons terms are
\begin{equation}
    \begin{split}
        \Omega_{3L} &= \text{Tr}\left( \boldsymbol{\omega} \wedge \boldsymbol{d \omega} + \frac{2}{3} \boldsymbol{\omega} \wedge \boldsymbol{\omega} \wedge \boldsymbol{\omega} \right)\\
        &= \boldsymbol{\omega}^a{}_c \wedge \boldsymbol{d \omega}^c{}_a + \frac{2}{3} \boldsymbol{\omega}^a{}_c \wedge \boldsymbol{\omega}^c{}_d \wedge \boldsymbol{\omega}^d{}_a
    \end{split}
\end{equation}
and
\begin{equation}
    \begin{split}
        \Omega_{3Y} &= \text{Tr}\left(\boldsymbol{A} \wedge \boldsymbol{dA} + \frac{2}{3} \boldsymbol{A} \wedge \boldsymbol{A} \wedge \boldsymbol{A} \right)\\
        &= \boldsymbol{A}^I \wedge \boldsymbol{d A}^I + \frac{2}{3} f_{IJK} \boldsymbol{A}^I \wedge \boldsymbol{A}^J \wedge \boldsymbol{A}^K
    \end{split}
\end{equation}
respectively, where $\boldsymbol{\omega}$ is the spin connection one-form and $\boldsymbol{A}$ is the Yang--Mills gauge potential one-form. The~trace $\text{Tr}$ is over both gauge and Lorentz group indices. This modification of the Kalb--Ramond field strength leads to a corresponding modification of the Bianchi identity it satisfies, which can be written as follows~\cite{Duncan:1992vz}:
\begin{equation}
    \boldsymbol{d \mathcal{H}} = \frac{\alpha^\prime}{8 \kappa} \text{Tr} \left(\boldsymbol{R} \wedge \boldsymbol{R} - \boldsymbol{F} \wedge \boldsymbol{F}\right),
    \label{bianchi_heterotic_form}
\end{equation}
where $\boldsymbol{F} = \boldsymbol{dA} + \boldsymbol{A} \wedge \boldsymbol{A}$ is the Yang--Mills field strength two-form and $\boldsymbol{R}^a{}_b = \boldsymbol{d\omega}^a{}_b + \boldsymbol{\omega}^a{}_c \wedge \boldsymbol{\omega}^c{}_b$ is the curvature two-form. The~modified, nonzero quantity on the right-hand side of the Bianchi identity is the so-called ``mixed (gauge and gravitational) quantum anomaly''~\cite{Alvarez-Gaume:1983ihn}.

We now consider the bosonic part of the heterotic effective string action up to order $\mathcal{O}(\alpha^\prime)$. We will consider the case where the dilaton varies slowly or has been stabilized (e.g.,~by means of an appropriate, string loop-induced potential) to some constant value $\Phi_0$, which, without~loss of generality, we may set to zero. The~effective string action  is then as follows:
\begin{equation}\label{act0}
    S_B = S_0 + \alpha^\prime S_1,
\end{equation}
where\footnote{We use a $(-++\dots +)$ signature for the metric. The~Riemann tensor is defined as $R^\rho{}{}_{\sigma\mu\nu} = \partial_\mu \Gamma^\rho_{\nu\sigma} - \partial_\nu \Gamma^\rho_{\mu\sigma} + \Gamma^\rho_{\mu\lambda}\Gamma^\lambda_{\nu\sigma} - \Gamma^\rho_{\nu\lambda}\Gamma^\lambda_{\mu\sigma}$, while the Ricci tensor is defined as $R_{\mu\nu} = R^\lambda{}_{\mu\lambda\nu}$.}
\begin{equation}\label{act1}
    S_0 = \int \left(\frac{1}{2\kappa^2} R - \frac{1}{6} \mathcal{H}_{\mu \nu \lambda} \mathcal{H}^{\mu \nu \lambda} \right) \sqrt{-g} \, d^D x,
\end{equation}
is the leading term and~\cite{Duncan:1992vz}
\begin{equation}\label{act2}
    \begin{split}
        S_1 &= \int \bigg\{ \frac{1}{16 \kappa^2} \left(R_{\mu \nu \lambda \rho} R^{\mu \nu \lambda \rho} + \lambda_1 R_{\mu \nu} R^{\mu \nu} +\lambda_2 R^2\right) + \\
        &\phantom{=} - \frac{1}{8} \left[\left(1-\frac{1}{3 \sqrt{3}}\right) \mathcal{H}_{\mu \nu \kappa} \mathcal{H}_{\lambda \rho}{}^\kappa R^{\mu \nu \lambda \rho} + \lambda_3 \mathcal{H}_{\mu \lambda \rho} \mathcal{H}_{\nu}{}^{\lambda \rho} R^{\mu \nu} + \lambda_4 \mathcal{H}_{\mu \nu \lambda} \mathcal{H}^{\mu \nu \lambda} R \right] + \\
        &\phantom{=}+ \frac{\kappa^2}{24} \left[\left(1-\frac{2}{9 \sqrt{3}}\right) \mathcal{H}^{\mu \nu \lambda} \mathcal{H}_{\mu}{}^{\rho \sigma} \mathcal{H}_{\nu \rho}{}^\kappa \mathcal{H}_{\lambda \sigma \kappa} + \lambda_5 \mathcal{H}^{\mu \lambda \rho} \mathcal{H}_{\nu \lambda \rho} \mathcal{H}^{\nu \kappa \sigma} \mathcal{H}_{\mu \kappa \sigma} + \lambda_6 (\mathcal{H}_{\mu \nu \lambda} \mathcal{H}^{\mu \nu \lambda})^2 \right] \\
        & \phantom{=}+ \frac{\lambda_7}{2} \nabla_\mu \mathcal{H}^{\mu \lambda \rho} \nabla^\nu \mathcal{H}_{\nu \lambda \rho} \bigg\}  \sqrt{-g} \, d^D x\,,
    \end{split}
\end{equation}
where $\nabla_\mu$ denotes the (torsion-free) gravitational covariant derivative, 
and $\alpha^\prime$ is the perturbative correction to the action
\footnote{There are two more terms of the form $\nabla \mathcal{H} \nabla \mathcal{H}$ that we may construct, namely $\nabla_{\gamma} \mathcal{H}_{\alpha\beta\delta} \nabla^{\delta} \mathcal{H}^{\alpha\beta\gamma}$ and $\nabla_{\delta} \mathcal{H}_{\alpha\beta\gamma} \nabla^{\delta} \mathcal{H}^{\alpha\beta\gamma}$. However, these can both be ``converted'' to $\nabla_\mu \mathcal{H}^{\mu \lambda \rho} \nabla^\nu \mathcal{H}_{\nu \lambda \rho}$ via partial integrations of modulo total derivative terms
	. Note that in the case of $\nabla_{\delta} \mathcal{H}_{\alpha\beta\gamma} \nabla^{\delta} \mathcal{H}^{\alpha\beta\gamma}$ especially, we need to make use of the Bianchi identity $\nabla_{[\delta} \mathcal{H}_{\alpha \beta \gamma]} = 0$ to achieve this (the reader should note at this stage that the $\mathcal O(\alpha^\prime)$ corrections, due to the GS mechanism in~the Bianchi identity  \eqref{bianchi_heterotic_form}, are ignored when the identity is implemented in the aforementioned  four-derivative terms
$\nabla \mathcal H \, \nabla \mathcal H$ as~a result of the truncation of the respective effective action to four spacetime derivatives, which we assume throughout this work).}.
The coefficients $\lambda_1, \ldots \lambda_7$ are not uniquely identified due to field redefinition ambiguities present in the theory, which leave the perturbative S-matrix invariant~\cite{Zwiebach:1985uq,Gross:1986mw,Metsaev:1987zx,Bento:1986hx,Bern:1987bq,Duncan:1992vz}. 

In what follows, we shall fix these redefinitions with~the aim of arriving at the most general string effective action to order $\mathcal O(\alpha^\prime)$ based on the graviton and antisymmetric tensor fields (having stabilized the dilaton). Motivated by the desire to apply it to phenomenologically realistic cosmologies, in~(3+1)-dimensional spacetimes after string compactification~\cite{Green_Schwarz_Witten_2012}, and~in particular the StRVM cosmology, we shall impose unitarity and a torsional interpretation of the totally antisymmetric tensor field strength, features that have been used in that approach~\cite{bms,bms2,ms1,ms2}. 

\section{Field Redefinitions and the Most General Unitary $\mathcal O(\alpha^\prime)$ String-Inspired Effective Action with Totally Antisymmetric~Torsion}
\label{sec:redef} 

In this section, we shall discuss the most general $\mathcal O(\alpha^\prime)$ 
low-energy string effective action, compatible with the $D(=10)$-dimensional 
string theory (although we perform the analysis in a general number of spacetime dimensions $D$, we have in mind nonetheless the low-energy theory corresponding to realistic superstring, and~in particular heterotic~\cite{Green_Schwarz_Witten_2012}, string theories). According to the equivalence theorem~\cite{Kamefuchi:1961sb,Salam:1970fso,Kallosh:1972ap,Bergere:1975tr,Georgi:1991ch,Weinberg:1995mt}, the~perturbative scattering (S-)matrices of quantum field theories linked through \emph{local} field redefinitions are the same.
This is what we make use of in our search for the most general effective action that describes StRVM embedded in realistic string theories, serving as consistent Ultra Violet (UV) completions of the model, compatible with quantum gravity. The~equivalence theorem is valid, and~thus consistent with the intuition that field redefinitions are a mere change in variables in a path integral, provided that the redefinitions are local, invertible (i.e.,~non singular), and do not affect the boundary conditions of the system, nor the spectrum of its asymptotic particle~states.

The S-matrix associated with the the effective action \eqref{act0} (\eqref{act1}, \eqref{act2}) remains invariant under the following transformations of the graviton $g_{\mu\nu}$ and Kalb--Ramond $B_{\mu\nu}$ fields:
\begin{align}\label{fred}
    g^\prime_{\mu\nu} &= g_{\mu\nu} + \alpha^\prime T_{\mu\nu}, \nonumber \\
    B^\prime_{\mu\nu} &= B_{\mu\nu} + \alpha^\prime F_{\mu\nu}
\end{align}
where
\begin{align}\label{fred2}
    T_{\mu\nu} &= B_1 R_{\mu\nu} + B_2 R g_{\mu\nu} + B_3 \kappa^2 \mathcal{H}_{\mu\alpha\beta} \mathcal{H}_\nu{}^{\alpha\beta} + B_4  \kappa^2 g_{\mu\nu} \mathcal{H}_{\alpha\beta\gamma} \mathcal{H}^{\alpha\beta\gamma}, \nonumber \\
    F_{\mu\nu} &= B_5 \nabla_\alpha \mathcal{H}^\alpha{}_{\mu\nu}.
\end{align}
It 
 is obvious that only the $\mathcal{O}(\alpha^\prime{}^0)$ terms yield $\mathcal{O}(\alpha^\prime{})$ terms under this transformation, while $\mathcal{O}(\alpha^\prime{})$ terms yield only higher-order $\mathcal{O}(\alpha^\prime{}^2)$ terms. The~transformed $\mathcal{O}(\alpha^\prime)$ action thus reads
\begin{equation*}
    S^\prime = S_0 + \alpha^\prime (\delta S_0 + S_1),
\end{equation*}
where $\delta S_0$ can be calculated to be
\begin{equation}
    \begin{split}
        \delta S_0 &= \frac{1}{2\kappa^2} \int \bigg(\frac{1}{2} RT - T^{\mu \nu} R_{\mu \nu} -\frac{\kappa^2}{6} \mathcal{H}_{\mu \nu \lambda} \mathcal{H}^{\mu \nu \lambda} T + \kappa^2 \mathcal{H}_{\mu \lambda \rho} \mathcal{H}_{\nu}{}^{\lambda \rho} T^{\mu \nu} + \\
        &+2 \kappa^2 F_{\nu \lambda} \nabla_\mu \mathcal{H}^{\mu \nu \lambda} \bigg) \sqrt{-g} \, d^D x \\
        &=\int \bigg[\frac{1}{4\kappa^2}\left(B_1 + (D-2)B_2\right) R^2 - \frac{B_1}{2 \kappa^2} R_{\mu \nu} R^{\mu \nu}  +\nonumber \\ &+ \frac{1}{4}\left(B_3 + (D-2)B_4 - \frac{B_1}{3} - \frac{(D-6)}{3} B_2 \right) \mathcal{H}_{\mu \nu \lambda} \mathcal{H}^{\mu \nu \lambda} R + \\
        &\phantom{= \int } + \frac{1}{2}(B_1 - B_3) \mathcal{H}_{\mu \lambda \rho} \mathcal{H}_{\nu}{}^{\lambda \rho} R^{\mu \nu} + \frac{\kappa^2}{2} B_3 \mathcal{H}^{\mu \lambda \rho} \mathcal{H}_{\nu \lambda \rho} \mathcal{H}^{\nu \kappa \sigma} \mathcal{H}_{\mu \kappa \sigma} - \nonumber \\ &- \frac{\kappa^2}{12}\left(B_3 + (D-6) B_4 \right) (\mathcal{H}_{\mu \nu \lambda} \mathcal{H}^{\mu \nu \lambda})^2  + B_5 \nabla_\mu \mathcal{H}^{\mu \lambda \rho} \nabla^\nu \mathcal{H}_{\nu \lambda \rho} \bigg] \sqrt{-g} \, d^D x.
    \end{split}
\end{equation}
This, in~turn, means that the $\mathcal{O}(\alpha^\prime)$ action becomes
\begin{equation}
    \begin{split}
        (\delta S_0 + S_1) &= \int \bigg\{ \frac{1}{16 \kappa^2} \left( R_{\mu \nu \lambda \rho} R^{\mu \nu \lambda \rho} + \underbrace{\left(\lambda_1 - 8 B_1\right)}_{\hat{\lambda}_1} R_{\mu \nu} R^{\mu \nu} + \underbrace{\left(\lambda_2 + 4 B_1 + 4(D-2) B_2 \right) R^2}_{\hat{\lambda}_2} \right) + \\
        &\phantom{=} - \frac{1}{8} \bigg[ \left(1-\frac{1}{3 \sqrt{3}}\right) \mathcal{H}_{\mu \nu \kappa} \mathcal{H}_{\lambda \rho}{}^\kappa R^{\mu \nu \lambda \rho} + \underbrace{\left(\lambda_3 -4 B_1 + 4 B_3\right)}_{\hat{\lambda}_3} \mathcal{H}_{\mu \lambda \rho} \mathcal{H}_{\nu}{}^{\lambda \rho} R^{\mu \nu} + \\
        &\phantom{=} + \underbrace{\left(\lambda_4 - 2 B_3 - 2 (D-2) B_4 + \frac{2}{3} B_1 + \frac{2 (D-6)}{3} B_2 \right)}_{\hat{\lambda}_4} \mathcal{H}_{\mu \nu \lambda} \mathcal{H}^{\mu \nu \lambda} R \bigg] + \\
        &\phantom{=}+ \frac{\kappa^2}{24} \bigg[ \left(1-\frac{2}{9 \sqrt{3}}\right) \mathcal{H}^{\mu \nu \lambda} \mathcal{H}_{\mu}{}^{\rho \sigma} \mathcal{H}_{\nu \rho}{}^\kappa \mathcal{H}_{\lambda \sigma \kappa} + \underbrace{\left(\lambda_5 + 12 B_3\right)}_{\hat{\lambda}_5} \mathcal{H}^{\mu \lambda \rho} \mathcal{H}_{\nu \lambda \rho} \mathcal{H}^{\nu \kappa \sigma} \mathcal{H}_{\mu \kappa \sigma} + \\
        &\phantom{=} + \underbrace{\left(\lambda_6 - 2 B_3 - 2 (D-6) B_4\right)}_{\hat{\lambda}_6} (\mathcal{H}_{\mu \nu \lambda} \mathcal{H}^{\mu \nu \lambda})^2 \bigg] + \nonumber \\ & + \underbrace{\left(\frac{\lambda_7}{2} + B_5\right)}_{\hat{\lambda}_7} \nabla_\mu \mathcal{H}^{\mu \lambda \rho} \nabla^\nu \mathcal{H}_{\nu \lambda \rho} \bigg\} \sqrt{-g} \, d^D x.
    \end{split}
\end{equation}
Therefore, the~seven coefficients $\lambda_1, \ldots, \lambda_7$ are specified within a five-parameter space spanned by $B_1,\ldots,B_5$. One particular set of coefficients that yields a unitary action that matches string amplitudes is given by $\lambda_1 = -4, \lambda_2 = 1, \lambda_3 = -2, \lambda_4 = \frac{1}{3},$  $\lambda_5 = -3, \lambda_6 = \frac{2}{3}, \lambda_7 = 0$ \cite{Duncan:1992vz}. Plugging these in, we get the most general $\mathcal{O}(\alpha^\prime)$ effective string action in $D$ spacetime dimensions:
\begin{equation}
    \begin{split}
        (\delta S_0 + S_1)_{\text{General}} &= \int \bigg\{ \frac{1}{16 \kappa^2} \left( R_{\mu \nu \lambda \rho} R^{\mu \nu \lambda \rho} - \left(8 B_1 + 4 \right) R_{\mu \nu} R^{\mu \nu} + \left( 1 + 4 B_1 + 4 (D-2) B_2 \right) R^2 \right) + \\
        &\phantom{=} - \frac{1}{8} \bigg[ \left(1-\frac{1}{3 \sqrt{3}}\right) \mathcal{H}_{\mu \nu \kappa} \mathcal{H}_{\lambda \rho}{}^\kappa R^{\mu \nu \lambda \rho} + \left( 4 B_3 - 4 B_1 - 2\right) \mathcal{H}_{\mu \lambda \rho} \mathcal{H}_{\nu}{}^{\lambda \rho} R^{\mu \nu} + \\
        &\phantom{= \quad \,\,} + \left( \frac{1}{3} + \frac{2}{3} B_1 + \frac{2(D-6)}{3} B_2 - 2 B_3 - 2 (D-2) B_4 \right) \mathcal{H}_{\mu \nu \lambda} \mathcal{H}^{\mu \nu \lambda} R \bigg] + \\
        &\phantom{=}+ \frac{\kappa^2}{24} \bigg[ \left(1-\frac{2}{9 \sqrt{3}}\right) \mathcal{H}^{\mu \nu \lambda} \mathcal{H}_{\mu}{}^{\rho \sigma} \mathcal{H}_{\nu \rho}{}^\kappa \mathcal{H}_{\lambda \sigma \kappa} + \left(12 B_3 - 3\right) \mathcal{H}^{\mu \lambda \rho} \mathcal{H}_{\nu \lambda \rho} \mathcal{H}^{\nu \kappa \sigma} \mathcal{H}_{\mu \kappa \sigma} + \\
        &\phantom{= \qquad} + \left(\frac{2}{3} - 2 B_3 - 2 (D-6) B_4\right) (\mathcal{H}_{\mu \nu \lambda} \mathcal{H}^{\mu \nu \lambda})^2 \bigg] + \nonumber \\ & + B_5 \nabla_\mu \mathcal{H}^{\mu \lambda \rho} \nabla^\nu \mathcal{H}_{\nu \lambda \rho} \bigg\}  \sqrt{-g} \, d^D x.
    \end{split}
\end{equation}
We may restrict these transformations such that the resulting action remains unitary by preserving the Gauss--Bonnet combination $(B_1=B_2=0)$ and removing the $\nabla_\mu \mathcal{H}^{\mu \lambda \rho} \nabla^\nu \mathcal{H}_{\nu \lambda \rho}$ term $(B_5=0)$. Thus, we obtain
\begin{equation}
    \begin{split}
        (\delta S_0 + S_1)_{\text{Unitary}} &= \int \bigg\{ \frac{1}{16 \kappa^2} \left( R_{\mu \nu \lambda \rho} R^{\mu \nu \lambda \rho} - 4 R_{\mu \nu} R^{\mu \nu} + R^2 \right) + \\
        &\phantom{=} - \frac{1}{8} \bigg[ \left(1-\frac{1}{3 \sqrt{3}}\right) \mathcal{H}_{\mu \nu \kappa} \mathcal{H}_{\lambda \rho}{}^\kappa R^{\mu \nu \lambda \rho} + \left( 4 B_3 - 2\right) \mathcal{H}_{\mu \lambda \rho} \mathcal{H}_{\nu}{}^{\lambda \rho} R^{\mu \nu} + \\
        &\phantom{= \quad \,\,} + \left( \frac{1}{3} - 2 B_3 - 2 (D-2) B_4 \right) \mathcal{H}_{\mu \nu \lambda} \mathcal{H}^{\mu \nu \lambda} R \bigg] + \\
        &\phantom{=}+ \frac{\kappa^2}{24} \bigg[ \left(1-\frac{2}{9 \sqrt{3}}\right) \mathcal{H}^{\mu \nu \lambda} \mathcal{H}_{\mu}{}^{\rho \sigma} \mathcal{H}_{\nu \rho}{}^\kappa \mathcal{H}_{\lambda \sigma \kappa} + \left(12 B_3 - 3\right) \mathcal{H}^{\mu \lambda \rho} \mathcal{H}_{\nu \lambda \rho} \mathcal{H}^{\nu \kappa \sigma} \mathcal{H}_{\mu \kappa \sigma} + \\
        &\phantom{= \qquad} + \left(\frac{2}{3} - 2 B_3 - 2 (D-6) B_4\right) (\mathcal{H}_{\mu \nu \lambda} \mathcal{H}^{\mu \nu \lambda})^2 \bigg] \bigg\}  \sqrt{-g} \, d^D x,
    \end{split}
    \label{unitary_string_action_general}
\end{equation}
which is the most general unitary heterotic string effective action that matches string amplitudes in $D$ dimensions.

We will now show (as has already been accomplished in~\cite{Bento:1986hx}, for example) that the condition of unitarity is incompatible with the Kalb--Ramond field strength, having the role of torsion in general $D$ dimensions. To~accomplish that, we need to categorize the quadratic curvature invariants in the generalized curvature scheme, i.e.,~in the case where $H_{\mu \nu \lambda}$ assumes the role of torsion. We thus define a contorted connection by
\begin{equation}
    \bar{\Gamma}^\lambda_{\mu\nu} = \Gamma^\lambda_{\mu\nu} + \frac{\kappa}{\sqrt{3}} \mathcal{H}^\lambda{}_{\mu \nu},
\end{equation}
such that the torsion tensor is given as $T_{\mu \nu \lambda} = \frac{2 \kappa}{\sqrt{3}} \mathcal{H}_{\mu \nu \lambda}$. In~this scheme, the~leading term in the effective action can be written as follows (see Appendix (\ref{sec:identities}) for more details):
\begin{equation}
    \begin{split}
        S_0 &= \int \left(\frac{1}{2\kappa^2} R - \frac{1}{6} \mathcal{H}_{\mu \nu \lambda} \mathcal{H}^{\mu \nu \lambda} \right) \sqrt{-g} \, d^D x \\
        &= \int \frac{1}{2 \kappa^2} \bar{R} \sqrt{-g} \, d^D x.
    \end{split}
\end{equation}
where $\bar{R}$ is the generalized Ricci scalar. In~quadratic order, there are six different scalar invariants we may construct, which correspond to the different ways one may contract the generalized Riemann tensor (which is less symmetric than the Levi--Civita connection Riemann tensor), namely as follows:
\begin{equation}
    \bar{R}_{\mu \nu \lambda \rho} \bar{R}^{\mu \nu \lambda \rho}, \quad \bar{R}_{\mu \nu \lambda \rho} \bar{R}^{\mu \lambda \nu \rho}, \quad \bar{R}_{\mu \nu \lambda \rho} \bar{R}^{\lambda \rho \mu \nu}, \quad \bar{R}_{\mu \nu} \bar{R}^{\mu \nu}, \quad \bar{R}_{\mu \nu} \bar{R}^{\nu \mu}, \quad \bar{R}^2.
\end{equation}
This ``basis'' of invariants is not unique, and~in fact, it is more convenient to work in a different basis, that of~\cite{Baekler:2011jt}, as follows:
\begin{align}
    \bar{G}^+_1 &= \bar{R}^2, \\
    \bar{G}^+_2 &= \bar{R}_{\mu \nu \lambda \rho} \bar{R}^{\mu \nu \lambda \rho}, \\
    \bar{G}^+_3 &= \bar{R}_{\mu \nu \lambda \rho} \bar{R}^{\lambda \rho \mu \nu}, \\
    \bar{G}^+_4 &= \bar{R}_{\mu \nu \lambda \rho} \bar{R}^{\lambda \rho \mu \nu} - 4 \bar{R}_{\mu \nu} \bar{R}^{\nu \mu} + \bar{R}^2, \\
    \bar{G}^+_5 &= \bar{R}_{\mu \nu \lambda \rho} \bar{R}^{\mu \nu \lambda \rho} - 4 \bar{R}_{\mu \nu} \bar{R}^{\mu \nu} + \bar{R}^2,\\
    \bar{G}^+_6 &= -4(\bar{R}_{\mu \nu \lambda \rho} \bar{R}^{\mu \nu \lambda \rho} - 4 \bar{R}_{\mu \nu \lambda \rho} \bar{R}^{\mu \lambda \nu \rho} + \bar{R}_{\mu \nu \lambda \rho} \bar{R}^{\lambda \rho \mu \nu}).
\end{align}
In this basis, the~two Gauss--Bonnet-like terms $\bar{G}^+_4, \bar{G}^+_5$ appear explicitly.  We may calculate all these scalar invariants (which are valid for a connection with general torsion) in our specific case, where the torsion is totally antisymmetric and the Bianchi identity $\nabla_{[\rho} \mathcal{H}_{\mu \nu \lambda]} = 0$  also holds\footnote{In our string context, the~anomaly terms in the Bianchi identity \eqref{bianchi_heterotic_form} contribute terms of higher order 
in $\alpha^\prime$ to the effective action; hence, they are ignored here.}. 
By doing that, we obtain modulo total derivatives if we assume that these terms are part of an action:
\begin{align}
    \bar{G}^+_1 = R^2 - \frac{2}{3} \kappa^2 \mathcal{H}_{\mu \nu \lambda} \mathcal{H}^{\mu \nu \lambda} R + \frac{1}{9} \kappa^4 (\mathcal{H}_{\mu \nu \lambda} \mathcal{H}^{\mu \nu \lambda})^2,
    \end{align}
\begin{align}
        \bar{G}^+_2 &= R_{\mu \nu \lambda \rho} R^{\mu \nu \lambda \rho} +\frac{2}{3} \kappa^2 \mathcal{H}_{\mu \nu \kappa} \mathcal{H}_{\lambda \rho}{}^\kappa R^{\mu \nu \lambda \rho} - \frac{4}{3} \kappa^2 \mathcal{H}_{\mu \lambda \rho} \mathcal{H}_{\nu}{}^{\lambda \rho} R^{\mu \nu} + \frac{4}{3} \kappa^2 (\nabla_\lambda H^{\mu \nu \lambda})(\nabla^\rho H_{\mu \nu \rho})\nonumber \\
        & - \frac{2}{9} \kappa^4 \mathcal{H}^{\mu \nu \lambda} \mathcal{H}_{\mu}{}^{\rho \sigma} \mathcal{H}_{\nu \rho}{}^\kappa \mathcal{H}_{\lambda \sigma \kappa} + \frac{2}{9} \kappa^4 \mathcal{H}^{\mu \lambda \rho} \mathcal{H}_{\nu \lambda \rho} \mathcal{H}^{\nu \kappa \sigma} \mathcal{H}_{\mu \kappa \sigma},
    \end{align}
\begin{align}
    \bar{G}^+_3 &= R_{\mu \nu \lambda \rho} R^{\mu \nu \lambda \rho} - 2 \kappa^2 \mathcal{H}_{\mu \nu \kappa} \mathcal{H}_{\lambda \rho}{}^\kappa R^{\mu \nu \lambda \rho} + \frac{4}{3} \kappa^2 \mathcal{H}_{\mu \lambda \rho} \mathcal{H}_{\nu}{}^{\lambda \rho} R^{\mu \nu} - \frac{4}{3} \kappa^2 (\nabla_\lambda H^{\mu \nu \lambda})(\nabla^\rho H_{\mu \nu \rho}) \nonumber \\
        & - \frac{2}{9} \kappa^4 \mathcal{H}^{\mu \nu \lambda} \mathcal{H}_{\mu}{}^{\rho \sigma} \mathcal{H}_{\nu \rho}{}^\kappa \mathcal{H}_{\lambda \sigma \kappa} + \frac{2}{9} \kappa^4 \mathcal{H}^{\mu \lambda \rho} \mathcal{H}_{\nu \lambda \rho} \mathcal{H}^{\nu \kappa \sigma} \mathcal{H}_{\mu \kappa \sigma},
    \end{align}
\begin{align}
        \bar{G}^+_4 &= R_{\mu \nu \lambda \rho} R^{\mu \nu \lambda \rho} - 4 R_{\mu \nu} R^{\mu \nu} + R^2 - 2 \kappa^2 \mathcal{H}_{\mu \nu \kappa} \mathcal{H}_{\lambda \rho}{}^\kappa R^{\mu \nu \lambda \rho} + 4 \kappa^2 \mathcal{H}_{\mu \lambda \rho} \mathcal{H}_{\nu}{}^{\lambda \rho} R^{\mu \nu} \nonumber \\ &- \frac{2}{3} \kappa^2 \mathcal{H}_{\mu \nu \lambda} \mathcal{H}^{\mu \nu \lambda} R- \frac{2}{9} \kappa^4 \mathcal{H}^{\mu \nu \lambda} \mathcal{H}_{\mu}{}^{\rho \sigma} \mathcal{H}_{\nu \rho}{}^\kappa \mathcal{H}_{\lambda \sigma \kappa} - \frac{2}{9} \kappa^4 \mathcal{H}^{\mu \lambda \rho} \mathcal{H}_{\nu \lambda \rho} \mathcal{H}^{\nu \kappa \sigma} \mathcal{H}_{\mu \kappa \sigma} \nonumber \\ &+ \frac{1}{9} \kappa^4 (\mathcal{H}_{\mu \nu \lambda} \mathcal{H}^{\mu \nu \lambda})^2,
    \end{align}
\begin{align}
        \bar{G}^+_5 &= R_{\mu \nu \lambda \rho} R^{\mu \nu \lambda \rho} - 4 R_{\mu \nu} R^{\mu \nu} + R^2 + \frac{2}{3} \kappa^2 \mathcal{H}_{\mu \nu \kappa} \mathcal{H}_{\lambda \rho}{}^\kappa R^{\mu \nu \lambda \rho} + \frac{4}{3} \kappa^2 \mathcal{H}_{\mu \lambda \rho} \mathcal{H}_{\nu}{}^{\lambda \rho} R^{\mu \nu} \nonumber \\ &- \frac{2}{3} \kappa^2 \mathcal{H}_{\mu \nu \lambda} \mathcal{H}^{\mu \nu \lambda} R- \frac{2}{9} \kappa^4 \mathcal{H}^{\mu \nu \lambda} \mathcal{H}_{\mu}{}^{\rho \sigma} \mathcal{H}_{\nu \rho}{}^\kappa \mathcal{H}_{\lambda \sigma \kappa} - \frac{2}{9} \kappa^4 \mathcal{H}^{\mu \lambda \rho} \mathcal{H}_{\nu \lambda \rho} \mathcal{H}^{\nu \kappa \sigma} \mathcal{H}_{\mu \kappa \sigma} \nonumber \\ &+ \frac{1}{9} \kappa^4 (\mathcal{H}_{\mu \nu \lambda} \mathcal{H}^{\mu \nu \lambda})^2,
    \end{align}
\begin{align}
    \bar{G}^+_6 &= \frac{64}{9} \kappa^4 \mathcal{H}^{\mu \nu \lambda} \mathcal{H}_{\mu}{}^{\rho \sigma} \mathcal{H}_{\nu \rho}{}^\kappa \mathcal{H}_{\lambda \sigma \kappa} - \frac{32}{9} \kappa^4 \mathcal{H}^{\mu \lambda \rho} \mathcal{H}_{\nu \lambda \rho} \mathcal{H}^{\nu \kappa \sigma} \mathcal{H}_{\mu \kappa \sigma}.
\end{align}
In this form, it becomes clear that $\bar{G}^+_1, \bar{G}^+_2, \bar{G}^+_3$ cannot be a part of a unitary action because~no linear combination eliminates the problematic terms for both the graviton and the Kalb--Ramond field\footnote{In the term $\bar{G}^+_1$, the problematic term is $R^2$. While this term does not spoil unitarity if replaced by a coupling to a scalar field (as is carried out in the Starobinsky model), this extra scalar degree of freedom is not present in the gravitational multiplet of the string, and we thus consider it unwanted in the context of the effective string action.}. Therefore, the~most general unitary action in the generalized curvature scheme (with $\mathcal H$ as torsion) is given by the linear combination $\frac{1}{\kappa^2}(A_4 \bar{G}^+_4 + A_5 \bar{G}^+_5 + A_6 \bar{G}^+_6)$, where $A_4, A_5, A_6$ are suitable constants:
{\small\begin{align}
        S_{\text{GCS}} &= \int \bigg\{\frac{1}{\kappa^2}(A_4+A_5)(R_{\mu \nu \lambda \rho} R^{\mu \nu \lambda \rho} - 4 R_{\mu \nu} R^{\mu \nu} + R^2) + \nonumber \\
        &\phantom{\int}+ \left(-2A_4+\frac{2}{3}A_5\right) \mathcal{H}_{\mu \nu \kappa} \mathcal{H}_{\lambda \rho}{}^\kappa R^{\mu \nu \lambda \rho} + \left(4 A_4 + \frac{4}{3} A_5 \right) \mathcal{H}_{\mu \lambda \rho} \mathcal{H}_{\nu}{}^{\lambda \rho} R^{\mu \nu} - \nonumber \\ &-\frac{2}{3} (A_4 + A_5) \mathcal{H}_{\mu \nu \lambda} \mathcal{H}^{\mu \nu \lambda} R + \frac{2}{9} \kappa^2 \left(32 A_6 - A_4 - A_5 \right) \mathcal{H}^{\mu \nu \lambda} \mathcal{H}_{\mu}{}^{\rho \sigma} \mathcal{H}_{\nu \rho}{}^\kappa \mathcal{H}_{\lambda \sigma \kappa} -  \nonumber \\ & -\frac{2}{9} \kappa^2 \left(16 A_6 + A_4 + A_5 \right) \mathcal{H}^{\mu \lambda \rho} \mathcal{H}_{\nu \lambda \rho} \mathcal{H}^{\nu \kappa \sigma} \mathcal{H}_{\mu \kappa \sigma} + \nonumber \\
        &\phantom{\int}+\frac{1}{9} \kappa^2 (A_4 + A_5) (\mathcal{H}_{\mu \nu \lambda} \mathcal{H}^{\mu \nu \lambda})^2 \bigg\}\sqrt{-g} \, d^D x.
\label{unitary_action_generalized_curvature_parametrized}
\end{align}}%
Our goal is to see if the unitary string action (\ref{unitary_string_action_general}) can be matched to this action. First of all, we can equate the terms not parametrized by $B_3, B_4$ and get the conditions
\begin{align}
    A_4 + A_5 &= \frac{1}{16}, \\
    -2A_4 + \frac{2}{3} A_5 &= - \frac{1}{8} \left(1- \frac{1}{3 \sqrt{3}}\right), \\
    \frac{2}{9} (32A_6 - A_4 - A_5) &= \frac{1}{24}\left(1-\frac{2}{9 \sqrt{3}}\right),
\end{align}
which may be solved to give
\begin{align}
    A_4 &= \frac{1}{16} - \frac{1}{64 \sqrt{3}}, \\
    A_5 &= \frac{1}{64 \sqrt{3}}, \\
    A_6 &= \frac{1}{64}\left(\frac{1}{2} - \frac{1}{12 \sqrt{3}}\right).
\end{align}
With this solution, the~action becomes
\begin{align}
        S_{\text{GCS}} &=\int\Bigg\{\frac{1}{16\kappa^{2}}\left(R_{\mu \nu \lambda \rho} R^{\mu \nu \lambda \rho}-4 R_{\mu \nu} R^{\mu \nu}+R^{2}\right) \nonumber \\
        &\phantom{\int} +\left(-\frac{1}{8}+\frac{1}{24\sqrt{3}}\right) \mathcal{H}_{\mu\nu\kappa} \mathcal{H}_{\lambda\rho}{}^{\kappa} R^{\mu\nu\lambda\rho}
        +\left(\frac{1}{4}-\frac{1}{24\sqrt{3}}\right) \mathcal{H}_{\mu\lambda\rho} \mathcal{H}_{\nu}{}^{\lambda\rho} R^{\mu\nu}
        -\frac{1}{24} \mathcal{H}_{\mu\nu\lambda} \mathcal{H}^{\mu\nu\lambda} R \nonumber \\
        &\phantom{\int} +\kappa^{2}\left[\frac{1}{24}\left(1-\frac{2}{9\sqrt{3}}\right)\right]
        \mathcal{H}^{\mu\nu\lambda} \mathcal{H}_{\mu}{}^{\rho\sigma} \mathcal{H}_{\nu\rho}{}^{\kappa} \mathcal{H}_{\lambda\sigma\kappa} +\kappa^{2}\left(-\frac{1}{24}+\frac{\sqrt{3}}{648}\right)
        \mathcal{H}^{\mu\lambda\rho} \mathcal{H}_{\nu\lambda\rho} \mathcal{H}^{\nu\kappa\sigma} \mathcal{H}_{\mu\kappa\sigma} \nonumber \\
        &\phantom{\int} + \kappa^{2}\left(\frac{1}{144}\right)( \mathcal{H}_{\mu\nu\lambda} \mathcal{H}^{\mu\nu\lambda})^{2}
        \Bigg\}\sqrt{-g} \, d^{D}x.
\label{unitary_action_generalized_curvature}
\end{align}
Comparison with (\ref{unitary_string_action_general}) shows that no choice of $B_3, B_4$ can yield (\ref{unitary_action_generalized_curvature}). (A particularly easy way to see this is to try to match both $\mathcal H_{\mu\lambda\rho} \mathcal H_{\nu}{}^{\lambda\rho} R^{\mu\nu}$ and $\mathcal{H}^{\mu\lambda\rho}\mathcal{H}_{\nu\lambda\rho}\mathcal{H}^{\nu\kappa\sigma}\mathcal{H}_{\mu\kappa\sigma}$. This, then, leads to contradicting solutions for the value of $B_3$). In~other words, in~$D > 4 $ dimensions, we have to choose between a unitary action and interpreting the Kalb--Ramond field strength as torsion~\cite{Bento:1986hx,Bern:1987bq}. However, this is \emph{not} necessarily the case in $D=4$. Here, due to over-antisymmetrization, the~following dimensionally dependent identities hold:
\begin{align}
    (\mathcal{H}_{\mu\nu\lambda} \mathcal{H}^{\mu\nu\lambda})^2 &= 6 \mathcal{H}^{\mu\nu\lambda} \mathcal{H}_{\mu}{}^{\rho\sigma} \mathcal{H}_{\nu\rho}{}^{\kappa} \mathcal{H}_{\lambda\sigma\kappa}, \\
    \mathcal{H}^{\mu\lambda\rho} \mathcal{H}_{\nu\lambda\rho} \mathcal{H}^{\nu\kappa\sigma} \mathcal{H}_{\mu\kappa\sigma} & = 2 \mathcal{H}^{\mu\nu\lambda} \mathcal{H}_{\mu}{}^{\rho\sigma} \mathcal{H}_{\nu\rho}{}^{\kappa} \mathcal{H}_{\lambda\sigma\kappa}, \\
    \mathcal{H}_{\mu\nu\kappa} \mathcal{H}_{\lambda\rho}{}^{\kappa} R^{\mu\nu\lambda\rho} & = 2 \mathcal{H}_{\mu\lambda\rho} \mathcal{H}_{\nu}{}^{\lambda\rho} R^{\mu\nu} - \frac{1}{3} \mathcal{H}_{\mu\nu\lambda} \mathcal{H}^{\mu\nu\lambda} R.
\end{align}
One way to confirm these identities is to replace $\mathcal{H}$ with its dual in $D=4$, i.e.,
\begin{align}\label{HdualV}
   \mathcal{H}^{\mu\nu\lambda} = \eta_{\mu \nu \lambda \rho} \mathcal V^\rho\,.  
\end{align}
We may use these identities to simplify the actions (\ref{unitary_string_action_general}) and (\ref{unitary_action_generalized_curvature_parametrized}) (post-compactification to (3+1)-dimensions) and
obtain the following\footnote{It should be noted that since the field redefinitions are performed in the $D=10$-dimensional low-energy heterotic string effective action, we should set $D=10$ in Equation~(\ref{unitary_string_action_general}). Then, we may proceed with the compactification to $D=4$ dimensions, and~only afterwards may we apply the dimensionally dependent identities to the (3+1)-dimensional action. Note that the processes of field redefinition and compactification do not ``commute'' in~the sense that if we compactify first and then carry out the field redefinitions in $D=4$, we obtain an action with different coefficients. For~more details, see Appendix (\ref{sec:beyondstrng}).}:
\begin{equation}
    \begin{split}
        (\delta S_0 + S_1)^{D=4}_{\text{Unitary}} &= \int \bigg\{ \frac{1}{16 \kappa^2} \left( R_{\mu \nu \lambda \rho} R^{\mu \nu \lambda \rho} - 4 R_{\mu \nu} R^{\mu \nu} + R^2 \right) + \\
        &\phantom{=} - \frac{1}{8} \bigg[ \left( 4 B_3 - \frac{2}{3 \sqrt{3}}\right) \mathcal{H}_{\mu \lambda \rho} \mathcal{H}_{\nu}{}^{\lambda \rho} R^{\mu \nu} + \left( \frac{1}{9\sqrt{3}} - 2 B_3 - 16 B_4 \right) \mathcal{H}_{\mu \nu \lambda} \mathcal{H}^{\mu \nu \lambda} R \bigg] + \\
        &\phantom{=}+ \frac{\kappa^2}{24} \bigg[ \left(-1-\frac{2}{9 \sqrt{3}} + 12 B_3 - 48 B_4 \right) \mathcal{H}^{\mu \nu \lambda} \mathcal{H}_{\mu}{}^{\rho \sigma} \mathcal{H}_{\nu \rho}{}^\kappa \mathcal{H}_{\lambda \sigma \kappa} \bigg] \bigg\}  \sqrt{-g} \, d^4 x,
    \end{split}
    \label{unitary_string_action_general_D=4}
\end{equation}
and\footnote{Interestingly, the~invariant $\bar{G}^+_6$ vanishes in $4D$ for a totally antisymmetric torsion. The~invariant $\bar{G}^+_4$, being the Gauss--Bonnet combination with torsion (Euler density in Einstein--Cartan spacetimes), reduces to the ordinary Gauss--Bonnet combination in the absence of torsion (Euler density in Einstein (general relativity) spacetimes).}
\begin{align}
        S_{\text{GCS}}^{D=4} &= \int \bigg\{\frac{1}{\kappa^2}(A_4+A_5)(R_{\mu \nu \lambda \rho} R^{\mu \nu \lambda \rho} - 4 R_{\mu \nu} R^{\mu \nu} + R^2) + \nonumber \\ &+ \frac{8}{3} A_5 \left( \mathcal{H}_{\mu \lambda \rho} \mathcal{H}_{\nu}{}^{\lambda \rho} R^{\mu \nu} - \frac{1}{3} \mathcal{H}_{\mu \nu \lambda} \mathcal{H}^{\mu \nu \lambda} R \right) \bigg\}\sqrt{-g} \, d^4 x.
\label{unitary_action_generalized_curvature_D=4}
\end{align}
We can then match the two actions by choosing
\begin{align}\label{AB}
    B_4 &= - \frac{1}{84} \left(\frac{1}{4} + \frac{5}{9\sqrt{3}}\right),\\
    B_3 &= \frac{1}{14}\left(1 - \frac{1}{9 \sqrt{3}} \right), \\
    A_4 &= \frac{1}{112} \left(\frac{17}{2} - \frac{11}{3 \sqrt{3}}\right), \\
    A_5 &= -\frac{3}{224}\left(1 - \frac{22}{9 \sqrt{3}}\right).
\end{align}
In this case, the~resulting unitary effective string action, where $\mathcal H$ is assumed to play the role of torsion, reads
\begin{equation}
    \begin{split}
        (\delta S_0 + S_1)^{D=4}_{\text{Unitary, Torsion}} &= \int \bigg\{ \frac{1}{16 \kappa^2} \left( R_{\mu \nu \lambda \rho} R^{\mu \nu \lambda \rho} - 4 R_{\mu \nu} R^{\mu \nu} + R^2 \right) + \\
        &\phantom{=} - \frac{1}{8} \left(\frac{2}{7} \left(1 - \frac{22}{9 \sqrt{3}} \right) \right) \bigg[ \mathcal{H}_{\mu \lambda \rho} \mathcal{H}_{\nu}{}^{\lambda \rho} R^{\mu \nu} -\frac{1}{3} \mathcal{H}_{\mu \nu \lambda} \mathcal{H}^{\mu \nu \lambda} R \bigg] \bigg\}  \sqrt{-g} \, d^4 x.
    \end{split}
\end{equation}

The full action can be written in terms of the generalized curvature as
\begin{align}\label{effstringactionintermsofgeneralizedcurvature}
        S &= \int \left[\frac{1}{2 \kappa^2} \bar{R} + \frac{\alpha^\prime}{\kappa^2} \left(A_4 \bar{G}^+_4 + A_5 \bar{G}^+_5\right) \right] \sqrt{-g} \, d^4 x \nonumber \\
        &= \int \Big[\frac{1}{2 \kappa^2} \bar{R} + \frac{\alpha^\prime}{\kappa^2} \Big[A_4 \left(\bar{R}_{\mu \nu \lambda \rho} \bar{R}^{\lambda \rho \mu \nu} - 4 \bar{R}_{\mu \nu} \bar{R}^{\nu \mu} + \bar{R}^2\right) +  \nonumber \\ &
      +  A_5 \left(\bar{R}_{\mu \nu \lambda \rho} \bar{R}^{\mu \nu \lambda \rho} - 4 \bar{R}_{\mu \nu} \bar{R}^{\mu \nu} + \bar{R}^2\right)\Big] \Big] \sqrt{-g} \, d^4 x \nonumber \\
        &= \int \bigg[\frac{1}{2 \kappa^2} \bar{R} + \frac{\alpha^\prime}{\kappa^2} \bigg[ \frac{1}{16}\bar{R}^2 + \frac{3}{224} \left(\frac{17}{3} - \frac{22}{9 \sqrt{3}}\right) \left(\bar{R}_{\mu \nu \lambda \rho} \bar{R}^{\lambda \rho \mu \nu} - 4 \bar{R}_{\mu \nu} \bar{R}^{\nu \mu}\right) \nonumber \\
        & - \frac{3}{224}\left(1 - \frac{22}{9 \sqrt{3}}\right) \left(\bar{R}_{\mu \nu \lambda \rho} \bar{R}^{\mu \nu \lambda \rho} - 4 \bar{R}_{\mu \nu} \bar{R}^{\mu \nu}\right)\bigg] \bigg] \sqrt{-g} \, d^4 x\,.
\end{align}

In $D=4$, the~Gauss--Bonnet combination is a total derivative, which we may ignore. Thus, the~full effective action reads
\begin{align}\label{string_effaction}
    S &= \int \Big[\frac{1}{2\kappa^2} R - \frac{1}{6} \mathcal{H}_{\mu \nu \lambda} \mathcal{H}^{\mu \nu \lambda} \nonumber \\ &+ \alpha^\prime \left(\frac{11}{126\sqrt{3}} - \frac{1}{28} \right) \left( \mathcal{H}_{\mu \lambda \rho} \mathcal{H}_{\nu}{}^{\lambda \rho} R^{\mu \nu} -\frac{1}{3} \mathcal{H}_{\mu \nu \lambda} \mathcal{H}^{\mu \nu \lambda} R\right) \Big] \sqrt{-g} \, d^4 x.
\end{align}

Since $\bar{G}^+_4$ is the Gauss--Bonnet combination in an Einstein--Cartan manifold, it is easy to see, from~\eqref{unitary_action_generalized_curvature_D=4} and \eqref{effstringactionintermsofgeneralizedcurvature},  that the effective string action is characterized by a single quadratic curvature invariant, namely $\bar{G}^+_5$, which provides the non-vanishing $\mathcal H \mathcal HR$ terms. In~the next section, we shall discuss the 
emergence of dynamical axion fields from the 
totally antisymmetric torsion interpretation of the Green--Schwarz-modified Kalb--Ramond field strength \eqref{KRH}, following~\cite{Duncan:1992vz} (or~\cite{bms} in the StRVM context).

\section{Axion-Like Fields from the $\mathcal H$-Torsion}\label{sec:Haxion}

To this end, we first rewrite the $D=4$ effective action \eqref{string_effaction} in a more convenient form, with~a generic coefficient of the higher-order corrections, as
\begin{equation}
    S = \int \left( \frac{1}{2\kappa^2} R - \frac{1}{6} \mathcal{H}_{\mu \nu \lambda} \mathcal{H}^{\mu \nu \lambda} + A \left( \mathcal{H}_{\mu}{}^{\lambda \rho} \mathcal{H}_{\nu \lambda \rho} R^{\mu \nu} - \frac{1}{3} \mathcal{H}_{\mu \nu \lambda} \mathcal{H}^{\mu \nu \lambda} R \right) \right)\sqrt{-g} \, d^4 x,
    \label{unitary_strin_action_final}
\end{equation}
where
\begin{align}\label{Adef}
    A = \alpha^\prime \left(\frac{11}{126 \sqrt{3}} - \frac{1}{28} \right) \simeq 0.015 \, \alpha^\prime \,,
\end{align}
in the heterotic string case. The reader should recall, from~the discussion in the previous section, that this value of the coefficient $A$ arose by taking the original action \eqref{act2} of~\cite{Duncan:1992vz}, with~coefficients matching the string scattering amplitudes~\cite{Gross:1986mw} and requiring a unitary action. Then, upon~performing field redefinitions and reduction (compactification) to $D=4$, one arrives at \eqref{unitary_string_action_general_D=4}. Finally, matching with the most general unitary action with a torsionful connection \eqref{unitary_action_generalized_curvature_D=4}, one obtains \eqref{unitary_strin_action_final}.

In $D=4$, the~Bianchi identity (\ref{bianchi_heterotic_form}) can be contracted with the Levi--Civita tensor $\eta_{\mu \nu \lambda \rho}$, so that it can be expressed, in~tensor notation, as~the scalar identity
\begin{equation}
    \begin{split}
        \eta_{\mu \nu \lambda \rho} \nabla^\mu \mathcal{H}^{\nu \lambda \rho} = &- \frac{\alpha^\prime}{16 \kappa} \left(R_{\mu \nu \lambda \rho} \tilde{R}^{\mu \nu \lambda \rho} - F_{\mu \nu} \tilde{F}^{\mu \nu} \right) \\
        &\equiv \mathcal{G}(\boldsymbol{\omega},\boldsymbol{A}).
    \end{split}
    \label{string_bianchi_index}
\end{equation}
The Levi--Civita tensor in $D=4$ is defined as~\cite{EGUCHI1980213}
\begin{equation}\label{levicivita}
    \eta_{\mu \nu \lambda \rho} = \sqrt{-g} \epsilon_{\mu \nu \lambda \rho}, \quad \eta^{\mu \nu \lambda \rho} = \frac{1}{\sqrt{-g}} \epsilon^{\mu \nu \lambda \rho} \,,
\end{equation}
where $\epsilon$ denotes the Levi--Civita symbol (tensor density) with the conventions $\epsilon_{0123}=+1$, ~$\epsilon^{0123}=-1,$ \emph{etc}. Furthermore, the~symbol $(\widetilde{\cdots})$ over the curvature and gauge field strength tensors denotes the corresponding duals, defined as
\begin{equation}\label{duals}
    \tilde{R}_{\mu \nu \lambda \rho} = \frac{1}{2} \eta_{\mu \nu \kappa \sigma} R^{\kappa \sigma}{}_{\lambda \rho}, \quad \tilde{F}_{\mu \nu} = \frac{1}{2} \eta_{\mu \nu \lambda \rho} F^{\lambda \rho}.
\end{equation}
We can now impose the Bianchi identity (\ref{string_bianchi_index}) as a constraint in the path integral of the action~(\ref{unitary_strin_action_final}), following~\cite{Duncan:1992vz,bms}. We accomplish this by inserting the identity as a delta functional in the path integral:
\begin{equation}\label{deltafunct}
    Z = \int \mathcal{D} g \mathcal{D} \mathcal{H} \delta(\eta^{\mu \nu \lambda \rho} \nabla_\mu \mathcal{H}_{\nu \lambda \rho} - \mathcal{G} (\boldsymbol{\omega}, \boldsymbol{A})) e^{i S}.
\end{equation}
We can express this delta functional as a path integral of a pseudoscalar Lagrange multiplier field $b(x)$\footnote{The partition function \eqref{deltafunct} is parity ($P$)-even, while the CS anomaly $\mathcal G (\boldsymbol{\omega}, \boldsymbol{A})$ is parity-odd. Hence, the~field $b(x)$ must be pseudoscalar so that the right-hand side of \eqref{deltab} below is parity-even, thus matching the even-parity nature of its left-hand side.}:
\begin{equation}\label{deltab}
    \delta(\eta^{\mu \nu \rho \sigma} \nabla_\mu \mathcal{H}_{\nu \rho \sigma} - \mathcal{G} (\boldsymbol{\omega}, \boldsymbol{A})) = \int \mathcal{D} b \, e^{-i \int \big[ \partial_\mu b(x) \eta^{\mu \nu \lambda \rho} \, \mathcal{H}_{\nu \lambda \rho} + b(x) \mathcal{G} (\boldsymbol{\omega}, \boldsymbol{A}) \big] \sqrt{-g} \, d^4 x},
\end{equation}
where we perform a partial integration to move the partial derivative to $b$ and discard the total derivative term. Thus, our effective action becomes
\begin{equation}
    \begin{split}
        S_{Eff} = \int &\bigg( \frac{1}{2\kappa^2} R - \frac{1}{6} \mathcal{H}_{\mu \nu \lambda} \mathcal{H}^{\mu \nu \lambda} + A \left( \mathcal{H}_{\mu}{}^{\lambda \rho} \mathcal{H}_{\nu \lambda \rho} R^{\mu \nu} - \frac{1}{3} \mathcal{H}_{\mu \nu \lambda} \mathcal{H}^{\mu \nu \lambda} R \right) \\
        & - \partial_\sigma b(x) \eta^{\sigma \mu \nu \lambda} \, \mathcal{H}_{\mu \nu \lambda} - b(x) \mathcal{G} (\boldsymbol{\omega}, \boldsymbol{A})\bigg)\sqrt{-g} \, d^4 x.
    \end{split}
\end{equation}
This action is still quadratic in $\mathcal{H}$ and, therefore, can be integrated out analytically. To~this end, we first write it in the following form:
\begin{equation}\label{seffbJ}
    S_{Eff} = \int \left( \frac{1}{2\kappa^2} R - \frac{1}{2} \mathcal{H}^{\alpha \beta \gamma} K^{\mu \nu \lambda}{}_{\alpha \beta \gamma} \mathcal{H}_{\mu \nu \lambda} + J^{\mu \nu \lambda}  \mathcal{H}_{\mu \nu \lambda} - b(x) \mathcal{G} (\boldsymbol{\omega}, \boldsymbol{A})\right)\sqrt{-g} \, d^4 x,
\end{equation}
where
\begin{equation}\label{Jdef}
    J^{\mu\nu\lambda} = - \eta^{\sigma\mu\nu\lambda}\, \partial_\sigma b = + \eta^{\mu\nu\lambda\sigma}\, \partial_\sigma b \,, 
\end{equation}
and
\begin{equation}
    K^{\mu \nu \lambda}{}_{\alpha \beta \gamma} = \frac{1}{3} \left(\delta^{[\mu}_{[\alpha} \delta^{\nu}_{\beta} \delta^{\lambda]}_{\gamma]} + A \left(2R \delta^{[\mu}_{[\alpha} \delta^\nu_\beta \delta^{\lambda]}_{\gamma]} - 6 R^{[\mu}{}_{[\alpha} \delta^\nu_\beta \delta^{\lambda]}_{\gamma]}\right) \right).
\end{equation}
The path integral then takes the (schematic) form
\begin{equation}
    Z = \int \mathcal{D} g \mathcal{D} b e^{i \int \left[ \frac{1}{2\kappa^2} R - b(x) \mathcal{G} (\boldsymbol{\omega}, \boldsymbol{A})\right]\sqrt{-g} \, d^4 x} \int \mathcal{D} \mathcal{H}_{\lambda \mu \nu} e^{i \int \left[-\frac{1}{2} \mathcal{H} K \mathcal{H} + J \mathcal{H} \right]\sqrt{-g} \, d^4 x},
\end{equation}
and we can proceed to integrate out $\mathcal{H}$. We perform a Wick rotation to a Euclidean version of the path integral as follows:
\begin{equation}
    Z_H = \int \mathcal{D} \mathcal{H} e^{\int \left(\left[\frac{1}{2} H K H - J H \right]\sqrt{-g} \, d^4 x\right)_E} = \mathcal{N} (\det K)^{-\frac{1}{2}} e^{- \int \left(\left[ \frac{1}{2} J K^{-1} J \right]\sqrt{-g} \, d^4 x\right)_E},
\end{equation}
where $\mathcal{N}$ is a normalization constant. We may define a tensor $Q^{\mu \nu \lambda}{}_{\alpha \beta \gamma}$ such that
\begin{equation}
    K^{\mu \nu \lambda}{}_{\alpha \beta \gamma} = \frac{1}{3} Q^{\mu \nu \lambda}{}_{\alpha \beta \gamma},
\end{equation}
So, $\det K = C \det Q$, where $C$ is a (formally infinite) field-independent ``constant'', with~no physical significance that can be absorbed in the path integral normalization constant $\mathcal{N}$, and~thus canceled when one computes normalized correlation functions that are physically~relevant. 

We can then write
\begin{equation}
    Z_H = \underbrace{\mathcal{N} C}_{=\Tilde{\mathcal{N}}}(\det Q)^{-\frac{1}{2}} e^{- \int \left(\left[ \frac{1}{2} J K^{-1} J \right]\sqrt{-g} \, d^4 x\right)_E},
\end{equation}
such that the overall factor of $\frac{1}{3}$ gets absorbed into the path integral normalization constant. Then, we can use the identity
\begin{equation}
    \det Q = e^{\ln{\det Q}} = e^{\text{Tr} \ln{Q}},
\end{equation}
where $\text{Tr}$ denotes the functional trace of an operator.
The computation follows standard treatments in quantum field theory~\cite{Srednicki:2007qs,dupuis,Weinberg:1995mt}, generalized here to covariant three-rank tensors. To~this end, we view $Q^{\mu \nu \lambda}{}_{\alpha \beta \gamma} (x)$ as the eigenfunction of a tensor operator $Q$ acting on the tensor field $\mathcal{H}_{\mu \nu \lambda}(x)$: 
\begin{equation}
    (Q \mathcal{H})_{\alpha \beta \gamma}(x) = Q^{\mu \nu \lambda}{}_{\alpha \beta \gamma}(x) \mathcal{H}_{\mu \nu \lambda} (x)\,.
\end{equation}
We can expand the operator $Q$ as
\begin{equation}
    Q = \int \sqrt{-g} \, d^4 z \, Q^{\mu \nu \lambda}{}_{\alpha \beta \gamma}(z) | z \rangle \langle z |,
\end{equation}
and any function of $Q$ is expanded as
\begin{equation}
    f(Q) = \int \sqrt{-g} \, d^4 z \, f(Q^{\mu \nu \lambda}{}_{\alpha \beta \gamma}(z)) | z \rangle \langle z |.
\end{equation}
The kernel is then defined as
\begin{equation}
    \langle x | f(Q) | y \rangle = \int \sqrt{-g} \, d^4 z \, f(Q^{\mu \nu \lambda}{}_{\alpha \beta \gamma}(z)) \langle x | z \rangle \langle z | y \rangle,
\end{equation}
and, since $\langle x | z \rangle = \delta^{(4)}(x-z)$, we obtain the following expression for the kernel:
\begin{equation}
    \langle x | f(Q) | y \rangle  = f(Q^{\mu \nu \lambda}{}_{\alpha \beta \gamma}(x)) \delta^{(4)}(x-y).
\end{equation}
The functional trace is then defined as
\begin{equation}
    \text{Tr} f(Q) = \int \sqrt{-g} \, d^4 x \, \text{tr} \langle x | f(Q) | x \rangle = \delta^{(4)}(0) \int \sqrt{-g} \, d^4 x \, \text{tr} f(Q^{\mu \nu \lambda}{}_{\alpha \beta \gamma}(x)),
\end{equation}
where the trace tr is over the internal (tensor) indices. In~our case, the~function $f$ is the natural logarithm and thus
\begin{equation}
    \text{Tr} \ln{Q} = \delta^{(4)}(0) \int \sqrt{-g} \, d^4 x \, \text{tr} \ln{(Q^{\mu \nu \lambda}{}_{\alpha \beta \gamma}(x))}.
\end{equation}
We can approximately calculate this expression by noticing that $Q$ is the identity tensor plus an $O(A) \sim O(a')$ correction:
\begin{equation}
    Q^{\mu \nu \lambda}{}_{\alpha \beta \gamma} = \bigg( \underbrace{\delta^{[\mu}_{[\alpha} \delta^{\nu}_{\beta} \delta^{\lambda]}_{\gamma]}}_{I^{\mu \nu \lambda}_{\alpha \beta \gamma}} + A \underbrace{\left(2R \delta^{[\mu}_{[\alpha} \delta^\nu_\beta \delta^{\lambda]}_{\gamma]} - 6 R^{[\mu}{}_{[\alpha} \delta^\nu_\beta \delta^{\lambda]}_{\gamma]}\right)}_{\Tilde{Q}^{\mu \nu \lambda}{}_{\alpha \beta \gamma}} \bigg).
\end{equation}
This allows us to expand $\ln{(Q^{\mu \nu \lambda}{}_{\alpha \beta \gamma}(x))}$ to order $O(A) \sim O(a')$ as
\begin{equation}
    \ln{(Q^{\mu \nu \lambda}{}_{\alpha \beta \gamma}(x))} = \ln{\left(I^{\mu \nu \lambda}_{\alpha \beta \gamma} + A \Tilde{Q}^{\mu \nu \lambda}{}_{\alpha \beta \gamma}\right)} \simeq A \Tilde{Q}^{\mu \nu \lambda}{}_{\alpha \beta \gamma}.
\end{equation}
On taking the trace, then, we obtain
\begin{equation}
    \text{tr} \ln{(Q^{\mu \nu \lambda}{}_{\alpha \beta \gamma}(x))} \simeq 2 A R.
\end{equation}
Thus, we calculate the (divergent) one-loop correction at order $O(A) \sim O(a')$ to be
\begin{equation}
    \text{Tr} \ln{Q} = \delta^{(4)}(0) \int 2 A R \sqrt{-g} \, d^4 x.
\end{equation}
Therefore, the~path integral takes the form

\begin{equation}
    \int \mathcal{D} \mathcal{H} e^{ \int \left[\frac{1}{2} \mathcal{H} K \mathcal{H} - J \mathcal{H} \right]\sqrt{-g} \, d^4 x} = e^{-\frac{1}{2} \text{Tr} \ln{K}} e^{- \int \left(\left[ \frac{1}{2} J K^{-1} J \right]\sqrt{-g} \, d^4 x\right)_E} = e^{- \int \left(\left[ \frac{1}{2} J K^{-1} J + \delta^{(4)}(0) A R\right]\sqrt{-g} \, d^4 x\right)_E}.
\end{equation}

We can then Wick rotate back to Minkowski space and write the full path integral as
\begin{equation}
    Z = \int \mathcal{D} g \mathcal{D} b e^{i \int \left[ \frac{1}{2\kappa^2} R - b(x) \mathcal{G} (\boldsymbol{\omega}, \boldsymbol{A}) + \frac{1}{2} J K^{-1} J + \delta^{(4)}(0) A R \right]\sqrt{-g} \, d^4 x}.
\end{equation}
Finally, we shall determine the inverse of $ K^{\mu \nu \lambda}{}_{\alpha \beta \gamma}$. Schematically, we have that
\begin{equation}
    K = \frac{1}{3}(I + A \Tilde{Q}),
\end{equation}
where
\begin{equation}
    I \equiv \delta^{[\mu}_{[\alpha} \delta^{\nu}_{\beta} \delta^{\lambda]}_{\gamma]}
\end{equation}
is the identity operator and
\begin{equation}\label{Qdef}
    \Tilde{Q} \equiv 2 A \left( R \delta^{[\mu}_{[\alpha} \delta^\nu_\beta \delta^{\lambda]}_{\gamma]} - 3 R^{[\mu}{}_{[\alpha} \delta^\nu_\beta \delta^{\lambda]}_{\gamma]}\right) 
\end{equation}
is a perturbation to the identity operator of order $\mathcal{O}(\alpha^\prime)$. Therefore (again, schematically), the~inverse is
\begin{equation}
    K^{-1}= \frac{3}{I + A\Tilde{Q}} \simeq 3 (I - A \Tilde{Q} + A^2 \Tilde{Q}^2 + \cdots).
\end{equation}
Since $A \propto a'$, we only keep the $\propto A$ term in the expansion. Therefore, we have the following order in $\alpha^\prime$:
\begin{equation}\label{Kinverse}
    (K^{-1})^{\mu \nu \lambda}{}_{\alpha \beta \gamma} \simeq 3 \left(\delta^{[\mu}_{[\alpha} \delta^{\nu}_{\beta} \delta^{\lambda]}_{\gamma]} - A \left(2R \delta^{[\mu}_{[\alpha} \delta^\nu_\beta \delta^{\lambda]}_{\gamma]} - 6 R^{[\mu}{}_{[\alpha} \delta^\nu_\beta \delta^{\lambda]}_{\gamma]}\right) \right) = 3 (I^{\mu \nu \lambda}_{\alpha \beta \gamma} - A \Tilde{Q}^{\mu \nu \lambda}{}_{\alpha \beta \gamma}).
\end{equation}

 Before proceeding with calculating $J K^{-1} J$, and~thus the resulting effective action in terms of metric and axion $b$ fields,
we should now give, for~completeness, the~modification of \eqref{dual} due to the higher-order corrections of the effective action \eqref{unitary_strin_action_final} proportional to the coefficient $A$ (\eqref{Adef}). This can be achieved by studying the saddle points of the corresponding path integral over the $\mathcal H$-field for the actions \eqref{seffbJ} and \eqref{Jdef}. The~latter corresponds to the classical equation of motion of the action with respect to the field strength $\mathcal H_{\mu\nu\lambda}$, which~reads
\begin{align}\label{Heq}
    \mathcal H_{\mu\nu\lambda} = ( K^{-1})_{\mu\nu\lambda}^{\quad \,\,\,\,\alpha\beta\gamma}\, J_{\alpha\beta\gamma}\,, 
\end{align}
where $J_{\alpha\beta\gamma}$ is given in \eqref{Jdef}, ~the inverse $( K^{-1})_{\mu\nu\lambda}^{\quad \,\, \,\,\alpha\beta\gamma}$
in \eqref{Kinverse} and \eqref{Qdef}, and the~linear order in $A \propto \alpha^\prime$ (cf. \eqref{Adef})
. After~some straightforward algebraic manipulations, we finally arrive at
{\small\begin{align}\label{Hbrel}
  \mathcal H_{\mu\nu\lambda} &= 3\, \Big( 1 - 2\, A\, R  \Big) \, J_{\mu\nu\lambda} + 18\, A\, R^\sigma_{\quad [\mu}\, J_{\nu\lambda]\sigma} + {\mathcal O}(A^2)  \nonumber \\
  & \stackrel{\eqref{Jdef}}{=} 3\Big(1 - 2\, A\, R \Big) \eta_{\mu\nu\lambda\rho}\, \partial^\rho b + 18\, A\, \, R^\sigma_{\,\,\,\,[\mu} \, \eta_{\nu\lambda]\sigma\rho}\, \partial^\rho b + {\mathcal O}(A^2)\,\nonumber \\
  &= 3 \eta_{\mu\nu\lambda\rho}\, \partial^\rho b + 6 A \left(3 R^\sigma_{\,\,\,\,[\mu} \, \eta_{\nu\lambda]\sigma\rho} - R \eta_{\mu\nu\lambda\rho} \right) \partial^\rho b + {\mathcal O}(A^2) \nonumber\\
  &= 3 \eta_{\mu\nu\lambda\rho}\, \partial^\rho b + 6 A \left(-\eta_{\rho \nu \lambda \sigma} R_\mu{}^\sigma + \eta_{\rho \mu \lambda \sigma} R_\nu{}^\sigma - \eta_{\rho \mu \nu \sigma} R_\lambda{}^\sigma - R \eta_{\mu\nu\lambda\rho} \right) \partial^\rho b + {\mathcal O}(A^2)\,,
\end{align}}%
Then, we can use the Schouten identity, namely the fact that over-antisymmetrized objects vanish, as follows:
\begin{equation}
    \eta_{[\mu \nu \lambda \sigma} R_{\rho]}{}^\sigma = \eta_{\mu \nu \lambda \sigma} R_{\rho}{}^\sigma - \eta_{\rho \nu \lambda \sigma} R_\mu{}^\sigma + \eta_{\rho \mu \lambda \sigma} R_\nu{}^\sigma - \eta_{\rho \mu \nu \sigma} R_\lambda{}^\sigma - \eta_{\mu \nu \lambda \rho} R = 0\,,
\end{equation}
to obtain
\begin{equation}\label{dual3}
    \mathcal H_{\mu\nu\lambda} = 3 \eta_{\mu\nu\lambda\rho}\, \partial^\rho b - 6 A \eta_{\mu \nu \lambda \sigma} R_{\rho}{}^\sigma \partial^\rho b + {\mathcal O}(A^2)\,,
\end{equation}
which is consistent with the generic result that a three-form in (3+1)-dimensional spacetimes is dual to a vector $\mathcal V^\mu$ in~ the sense of the relation
\begin{align}\label{dual2}
  \mathcal H_{\mu\nu\lambda} = \eta_{\mu\nu\lambda\sigma}\, \mathcal V^\sigma \Rightarrow \mathcal{V}^\kappa = - \frac{1}{6} \eta^{\mu \nu \lambda \kappa} \mathcal{H}_{\mu \nu \lambda} \,. 
\end{align}
Indeed, from~\eqref{Hbrel} and \eqref{dual2}, we can calculate the dual vector as
\begin{equation}
    \mathcal{V}^\kappa = 3 \partial^\kappa b - 6 A R_\rho{}^\kappa \partial^\rho b\,,
\end{equation}
and, as~a consistency check, we observe that this properly reproduces $\mathcal{H}$ when its dual is~taken.

For $A=0$, this yields \eqref{dual}, while for $A\ne 0$, we observe that one obtains modifications of the duality relation, which amount to a direction-dependent term proportional to components of the Ricci tensor contracted with $\partial^\mu b$.
In the context of our StRVM, we are interested in (3+1)-dimensional Einstein spaces, for~which $R_{\mu\nu} = \frac{R}{4}\, g_{\mu\nu}$, ~which \eqref{dual3} reduces to
\begin{align}\label{einsteinH}
\mathcal H^{\rm Einstein}_{\mu\nu\rho} =  3 \left( 1 - \frac{1}{2} \, A R \right) \eta_{\mu\nu\rho\sigma} \, \partial^\sigma b
+ \mathcal O(A^2)
\,,
\end{align}
This amounts to a modification of \eqref{dual} by a global rescaling factor depending on the torsion-free scalar curvature of~the maximally symmetric Einstein spacetime.
As we shall discuss later on in this article, the~quantity $| A\, R | \ll 1 $ during the RVM inflationary scenario, so any singular behavior is avoided in this~case.

After this digression, we can now proceed to calculate $J K^{-1} J$, which is a necessary step for obtaining the effective action in terms of the axion field. We have that
\begin{equation}
    \begin{split}
        J^{\alpha \beta \gamma} (K^{-1})^{\lambda \mu \nu}{}_{\alpha \beta \gamma} J_{\lambda \mu \nu} &= 3 J^{\alpha \beta \gamma} (J_{\alpha \beta \gamma} - A(2 R J_{\alpha \beta \gamma} - 6 R_\alpha{}^\lambda J_{\lambda \beta \gamma})) \\
        &= 3 J^{\alpha \beta \gamma} J_{\alpha \beta \gamma} - 6 A (R J^{\alpha \beta \gamma} J_{\alpha \beta \gamma} - 3 R_{\alpha}{}^\lambda J^{\alpha \beta \gamma} J_{\lambda \beta \gamma}).
    \end{split}
\end{equation}
We calculate
\begin{equation}
    J^{\alpha \beta \gamma} J_{\lambda \beta \gamma} = \partial_\rho b \partial^\sigma  b \eta^{\rho \alpha \beta \gamma} \eta_{\sigma \lambda \beta \gamma} = - 2 \partial_\rho b \partial^\sigma b (\delta^\rho_\sigma \delta^\alpha_\lambda - \delta^\rho_\lambda \delta^\alpha_\sigma) = 2\partial_\lambda b \partial^\alpha b - 2\partial_\rho b \partial^\rho b \delta^\alpha_\lambda,
\end{equation}
and, by~extension,
\begin{equation}
    J^{\alpha \beta \gamma} J_{\alpha \beta \gamma} = -6 \partial_\rho b \partial^\rho b.
\end{equation}
Therefore, we  finally have
\begin{equation}
    \frac{1}{2} J^{\alpha \beta \gamma} (K^{-1})^{\lambda \mu \nu}{}_{\alpha \beta \gamma} J_{\lambda \mu \nu} = - 9 \partial_\mu b \partial^\mu b + 18 A \partial_\mu b \partial_\nu b R^{\mu \nu}\,.
\end{equation}
The resulting effective action is therefore given by
\begin{equation}
    S_E = \int \left(\left(\frac{1}{2\kappa^2} + \delta^{(4)}(0) A \right) R - 9 \partial_\mu b \partial^\mu b + 18 A \partial_\mu b \partial_\nu b R^{\mu \nu} - b \mathcal{G} (\boldsymbol{\omega}, \boldsymbol{A})\right)\sqrt{-g} \, d^4 x.
\end{equation}
Upon rescaling $b \rightarrow \frac{1}{3 \sqrt{2}} b$, we finally obtain an effective action with canonical kinetic terms for the KR axion field\footnote{We follow here the conventions and normalizations of~\cite{Duncan:1992vz}, which differ from those of~\cite{bms} by a rescaling of $\alpha^\prime$ by a factor of $\sqrt{3}$.}:

\begin{align}
        S_E &= \int \bigg[ \left(\frac{1}{2\kappa^2} + \delta^{(4)}(0) 
        \alpha^\prime \left(\frac{11}{126\sqrt{3}} - \frac{1}{28} \right) \right) R \nonumber - \frac{1}{2} \partial_\mu b \partial^\mu b + \frac{\alpha^\prime}{\kappa}\frac{\sqrt{2}}{96} b \left(R_{\mu \nu \lambda \rho} \tilde{R}^{\mu \nu \lambda \rho} - F_{\mu \nu} \tilde{F}^{\mu \nu} \right) + \\
        &\phantom{= \int} +         \alpha^\prime \left(\frac{11}{126 \sqrt{3}} - \frac{1}{28} \right) \partial_\mu b \partial_\nu b R^{\mu \nu} \bigg]\sqrt{-g} \, d^4 x.
    \label{sea}
\end{align}

On replacing $\delta^{(4)}(0) \sim\Lambda^4$, where 
$\Lambda$ is a UV cutoff energy scale, we observe that 
the coefficient of the Einstein--Hilbert (scalar curvature) term in the effective action \eqref{sea} is {\it positive},~so unitarity is~guaranteed.  

In string effective field theories, the~UV cutoff is the string mass scale
\begin{align}\label{LMs}
 \Lambda=M_s = (\alpha^\prime)^{-1/2}\,, 
\end{align}
whose value is  bounded from above by $M_{\rm Pl}$~\cite{Green_Schwarz_Witten_2012}. Using \eqref{LMs}, the~final form of the string-inspired effective action \eqref{sea}, containing terms of the fourth order in derivatives and~respecting the unitarity and torsion interpretation of $H_{\mu\nu\rho}$, reads

\begin{align}
        S_E &= \int \bigg[ \frac{1}{2\kappa^2} \left( 1 + \frac{\kappa^2}{\alpha^\prime }\left(\frac{11}{63 \sqrt{3}} - \frac{1}{14} \right) \right) R \nonumber - \frac{1}{2} \partial_\mu b \partial^\mu b + \frac{\alpha^\prime}{\kappa}\frac{\sqrt{2}}{96} b \left(R_{\mu \nu \lambda \rho} \tilde{R}^{\mu \nu \lambda \rho} - F_{\mu \nu} \tilde{F}^{\mu \nu} \right) + \\
        &\phantom{= \int} - \frac{\alpha^\prime}{2}\, \left(\frac{1}{14} - \frac{11}{63 \sqrt{3}} \right) \partial_\mu b \partial_\nu b R^{\mu \nu} \bigg]\sqrt{-g} \, d^4 x.
    \label{sea2}
\end{align}

Therefore, we observe that the effects of keeping the complete four-derivative $\mathcal{O}(\alpha^\prime)$ terms in the effective action of the bosonic massless string multiplet, under~the assumption of constant dilaton, are the following:
\begin{itemize}
    \item Renormalization of the coefficient of the Ricci scalar R, and~thus Newton's constant. The~effective Newton's constant is smaller than the one before the $\mathcal O(\alpha^\prime)$ corrections are taken into account.
    \item Introduction of a non-minimal derivative coupling of the KR axion b with the Ricci tensor, which maintains the shift symmetry of the axion, as~it should, given that such terms originate from a path integration of the $H$-field strength, which is classically related to $b$ via \eqref{dual}.
\end{itemize}

It is worth noting that when expressed in terms of the dimensionless $\widetilde b(x) \equiv \kappa \, b(x)$ axion field, the~coefficients of the four-spacetime derivative terms, that is, the~CS anomalous terms and the non-minimal derivative coupling term, are both proportional to order $\alpha^\prime /\kappa^2 \lesssim 1$. However, the~relative magnitude of these terms can only be determined once we know the corresponding contributions at specific cosmological eras, which we shall carry out in the next section. Since we are working in an EFT/perturbative approach, the~ghosts introduced by both the CS and the non-minimal derivative coupling term can be ignored as long as we stay below the UV cutoff~\cite{Dorlis:2024yqw,Dorlis:2024uei}. We now proceed to calculate the equations of motion. It will be convenient in what follows to use the following generic form of our effective action \eqref{sea2}:
\begin{equation}\label{seaE}
    S_E = \int \bigg[ \frac{1}{2 q^2} R - \frac{1}{2} \partial_\mu b \partial^\mu b + \frac{\epsilon}{8} b R_{\mu \nu \lambda \rho} \tilde{R}^{\mu \nu \lambda \rho} - \frac{\lambda}{2} \partial_\mu b \partial_\nu b R^{\mu \nu} \bigg]\sqrt{-g} \, d^4 x,
\end{equation}
with
\begin{align}\label{defeps}
    \lambda &= \alpha^\prime \left(\frac{1}{14} - \frac{11}{63 \sqrt{3}} \right) < 0 \,, \nonumber \\
    \frac{1}{2q^2} &= \frac{1}{2\kappa^2} \left[ 1 - \frac{\kappa^2}{\alpha^\prime } \frac{\lambda}{\alpha^\prime} \right] \, , \nonumber \\
    \epsilon & = \frac{\alpha^\prime}{\kappa} \frac{\sqrt{2}}{12}. 
\end{align}

For completeness we note that on~using the Ricci-tensor identity 
$$ R_{\mu\nu} \, \nabla^\nu b = \Big[\nabla_\mu, \, \nabla_\nu \Big] \nabla^\nu b \,,$$
and integrating by parts in the last (the $\lambda$-dependent ) term on the right-hand of \eqref{seaE}, 
one can show that
this higher-order term can be written in the following form:
\begin{align}\label{alternateRicci}
    \int \nabla_\mu b \nabla_\nu b R^{\mu \nu} \sqrt{-g} \, d^4 x&= \underbrace{\int \nabla_\mu (\nabla_\nu b \nabla^\mu \nabla^\nu b - \nabla^\mu b \Box b)\sqrt{-g} \, d^4 x}_{=\rm Boundary \ term} + \nonumber \\ &+ \int \left[(\Box b)^2 - (\nabla_\mu \nabla_\nu b) (\nabla^\mu \nabla^\nu b)\right] \sqrt{-g} \, d^4 x,
\end{align}
where $\Box = \nabla_\rho \nabla^\rho$ is the (torsion-free) gravitational covariant of d'Alembertian, and~the boundary terms are dropped, assuming standard boundary conditions for fields in this Universe.

It is also important to notice that as~a result of this $\lambda$-dependent term,  the~kinetic terms of the axion field $b(x)$ in the effective action \eqref{seaE}
appear to 
correspond to a $\mathcal O(\alpha^\prime)$ ``corrected metric tensor''
\begin{align}\label{friedan}
g_{\mu\nu}^\prime (x) = g_{\mu\nu}(x) + \lambda \,  R_{\mu\nu}(x)\,,  \quad \lambda = \mathcal O(\alpha^\prime) \, ~~\quad ({\emph cf.}~ \eqref{defeps})\,,
\end{align}
of the type appearing in the seminal papers of Friedan~\cite{Friedan:1980jf,Friedan:1980jm}, which constitutes the basis for the conformal $\sigma$-model approach to target-space string effective actions~\cite{Metsaev:1987zx} and is perturbatively equivalent to the $S$-matrix approach~\cite{Gross:1986mw}.

Variation in the effective action \eqref{seaE} with respect to the metric gives equations of the following form:
\begin{equation}\label{graveom}
    \frac{1}{2 q^2} G_{\mu \nu} = \frac{\epsilon}{2} \mathcal{C}_{\mu \nu} + \frac{\lambda}{2} \Theta_{\mu \nu} + \frac{1}{2} {T_{\rm kin}^b}_{\mu \nu},
\end{equation}
where $G_{\mu \nu}$ is the Einstein tensor, ${T_{\rm kin}^b}_{\mu \nu}$ is the axion stress energy tensor (corresponding to the axion kinetic term only in \eqref{seaE})
\begin{align}\label{Tb}
 {T_{\rm kin}^b}_{\mu\nu} = \partial_\mu b\, \partial_\nu b - \frac{1}{2}\, g_{\mu\nu} \partial_\alpha b \, \partial^\alpha b \, ,
\end{align}
$
\mathcal{C_{\mu \nu}}$ is a variant of the  Cotton tensor (or more precisely the C-tensor)~\cite{Jackiw:2003pm,Alexander:2009tp,Duncan:1992vz}, given as
\begin{equation}\label{cotton}
    \mathcal{C}^{\mu\nu} = \nabla_\sigma \Big( \nabla_\rho b \, \Tilde{R}^{\rho(\mu \nu)\sigma} \Big)\,,
\end{equation}
and $\Theta_{\mu \nu}$ is the tensor that arises after variation in $\partial_\mu b \partial_\nu b R^{\mu \nu}$, given as
\begin{align}\label{Thetamn}
    \Theta_{\mu \nu} &= \nabla_\sigma \big((\nabla_\mu\nabla_\nu b) \nabla^\sigma b\big) - 2 R_{\sigma(\mu} (\nabla_{\nu)} b)(\nabla^\sigma b) -  \nonumber \\ &-g_{\mu\nu}\Big[(\nabla^\sigma b)\nabla_\sigma\Box b+\frac12(\Box b)^2+\frac12(\nabla_\sigma\nabla_\rho b)(\nabla^\sigma\nabla^\rho b)\Big].
\end{align}
The parameter $q$ in \eqref{defeps} plays the role of the reduced Planck constant, renormalized by the $\alpha^\prime$ corrections. So, from now on, we replace the following in all subsequent analyses:
\begin{align}\label{qk}
    q \to \kappa \,.
\end{align}
Variation with respect to $b$ gives the following scalar equation of motion:
\begin{equation}\label{scalareom}
    \nabla_\mu\left(\partial^\mu b + \lambda R^{\mu\nu}\partial_\nu b + \frac{\epsilon}{8} \mathcal{K}^\mu\right) = 0,
\end{equation}
with $\mathcal{K}^\mu$ being the gravitational Chern--Simons current, as follows:
\begin{align}\label{current}
    \nabla_\mu \mathcal{K}^\mu=R_{\alpha\beta\gamma\delta}\tilde R^{\alpha\beta\gamma\delta}.
\end{align}
We next proceed to discuss the effects of the higher-order term \eqref{alternateRicci} on the running-vacuum GW-condensate inflationary scenario of~\cite{bms,Dorlis:2024yqw}.

\section{Stringy Running Vacuum Model~Inflation}
\label{sec:strvminfl} 

Having arrived at the effective action \eqref{seaE} to the fourth-derivative order, we are now well equipped to revisit the stringy running vacuum model cosmology (StRVM)~\cite{bms0,bms,bms2,ms1,ms2} and~see how the higher-order terms affect the relevant conclusions on the physics of the associated inflation. To~this end, we shall first demonstrate that inflation is an admissible solution of the pertinent cosmological equations of motion, as~was the case for the initial StRVM. 
The scalar equation of motion (\ref{scalareom}) takes the form of a conserved current equation as follows:
\begin{equation}\label{CScurrent}
    \nabla_\mu J^\mu = 0\,, \qquad J^\mu = \partial^\mu b + \lambda R^{\mu\nu}\partial_\nu b + \frac{\epsilon}{8} \mathcal{K}^\mu\,.
\end{equation}
Therefore, we obtain
\begin{equation}
    \nabla_\mu J^\mu = \frac{1}{\sqrt{-g}} \partial_\mu (\sqrt{-g} J^\mu) = 0 \Rightarrow \partial_\mu (\sqrt{-g} J^\mu) = 0
    \label{scalar_eom}
\end{equation}
Upon assuming a spatially flat background of an expanding isotropic and homogeneous Friedman--Lemaitre--Robertson--Walker (FLRW) Universe, with~scale factor $a(t)$, in~which the 
axion-like fields depend only on cosmic time $t$, $b=b(t)$, the~expression for the current simplifies to
\begin{equation}
    J^\mu = \partial^\mu b + \lambda R^{\mu 0}\partial_0 b + \frac{\epsilon}{8} \mathcal{K}^\mu
\end{equation}
which means that (\ref{scalar_eom}) becomes
\begin{equation}\label{conseqJ}
   0 =  \partial_0 (\sqrt{-g} J^0) + \partial_i (\sqrt{-g} J^i) = \partial_0 (\sqrt{-g} J^0)
\end{equation}
where $J^i = \lambda R^{i 0}\partial_0 \, b + \frac{\epsilon}{8} \mathcal{K}^i$, and~the last equality stems from the assumption of 
isotropy and homogeneity in the Universe~\cite{bms}, which implies
\begin{align}\label{pariJi}
\partial_i \Big(\sqrt{-g}\, J^i\Big)=0\,.
\end{align}
 Solving Equation~(\ref{conseqJ}), we get
\begin{equation}
    J^0 = \frac{C}{\sqrt{-g}},
\end{equation}
where $C$ is a constant. Hence, we obtain
\begin{equation}
    J^0 = \partial^0 b + \lambda R^{00} \partial_0 b + \frac{\epsilon}{8} \mathcal{K}^0 = - \dot{b} + \lambda R^{00} \dot{b} + \frac{\epsilon}{8} \mathcal{K}^0 = \frac{C}{\sqrt{-g}},
\end{equation}
because $\partial_0 b = \dot{b}$ and $\partial^0 b = - \dot{b}$ in our metric signature convention. So, we have
\begin{equation}
    \dot{b} = - \frac{C}{\sqrt{-g} \left(1-\lambda R^{00}\right) } + \frac{\epsilon}{8\left(1-\lambda R^{00}\right)} \mathcal{K}^0
\end{equation}
The constant $C$ can be set to $0$ if one considers an inflationary spacetime~\cite{bms}. Furthermore, on~an FLRW background, we have that during inflation 
$a \sim \exp (H\, t)$, with~$H = H_I \simeq constant$ representing the approximately constant Hubble parameter
, 
for which the cosmological data imply the following~\cite{Planck:2018vyg}:
\begin{align}\label{HPl}
H = H_I \simeq {\rm constant} \, \lesssim \, 10^{-5} \, \kappa^{-1}. 
\end{align}

Thus, the~time--time component of the Ricci tensor equals approximately
\begin{align}\label{R00H}
R^{00} = - 3 \frac{\ddot{a}}{a} \simeq - 3 H^2\,,
\end{align} 
and thus
\begin{equation}\label{bdotK0}
    \dot{b} = \frac{\epsilon}{8\left(1 + 3 \lambda \, H^2 \right)} \mathcal{K}^0
\end{equation}

In the presence of primordial GW perturbations, the gravitational CS (gCS) anomaly term in \eqref{seaE} can condense at the end of the axion-dominated (preinflationary) stiff era~\cite{bms}. This can drive the Universe to an inflationary phase, characterized  by approximately constant gCS-anomaly condensate and Hubble parameter $H$, as well as~the linear axion $b$ potential in~\eqref{seaE}, breaking the axion shift symmetry. This leads to a metastable inflation~\cite{bms,Dorlis:2024yqw,Dorlis:2024uei}, due to the existence of imaginary parts in the condensate~\cite{Dorlis:2024uei}, which leads to a finite-lifetime inflationary era, and~eventual exit from it. However, 
although from~a purely dynamical system point of view~\cite{Dorlis:2024yqw,Dorlis:2025gvb}, it appears that it is the $b$-field that drives inflation, this is deceptive. The~condensate itself 
is a non-linear function of the (slowly varying) Hubble parameter $H$, depending on the fourth (and even higher)-order powers of it. In~this sense, it is such non-linearities of the gravitational sector that drive an inflation of the RVM type~\mbox{\cite{Sola:2007sv,Sola:2015rra,SolaPeracaula:2022hpd,SolaPeracaula:2025yco,SolaPeracaula:2026trz}}\footnote{For completeness, we mention that the inflationary exit in such cosmologies, which in general do not require external  inflaton fields, is characterized by prolonged reheating periods~\cite{Lima:2013dmf,Perico:2013mna}, and~perhaps early-matter dominance epochs, preceeding standard radiation. This may affect primordial black hole populations and thus gravitational wave profiles during the early radiation era, with~potentially detectable signatures in future intrerferometers~\cite{Mavromatos:2022yql,Tzerefos:2024rgb}.}.

To estimate the gCS condensation, the~authors of~\cite{Dorlis:2024yqw}, whose approach we follow here, considered tensor perturbations in the context of the effective gravitational Lagrangian~\eqref{seaE}, with~$\lambda=0$. They quantize them within an effective weak gravitational  
field theory action~\cite{Donoghue:1994dn,tHooft:1974toh}, following a canonical quantization approach to estimate the condensate, to~a leading (linear) order in a perturbative expansion in the small parameter $\dot b H^2 \ll 1\,.$

In the presence of the $\lambda$-dependent correction terms \eqref{alternateRicci} in \eqref{seaE}, higher-order corrections proportional to $(\dot b)^2$ are expected in the condensate, which will be shown below to be subleading. For~the $\lambda=0$ case, as~argued in~\cite{bms}, and~also confirmed by the dynamical system analysis of the StRVM inflation of~\cite{Dorlis:2024yqw}, as well as reviewed below, 
when there is a gravitational anomaly condensate, 
a spontaneous-Lorentz-violating solution of the $b$-axion equations of motion exists in which $\langle \mathcal K^0 \rangle \simeq {\rm constant}$.
We parametrize such constant quantity by~\cite{bms,Dorlis:2024yqw}
\begin{align}\label{Hbdot}
\dot{b} = (2\, \varepsilon)^{1/2} \, \kappa^{-1} \, H\,,
\end{align}
during inflation, 
where $\varepsilon$ (\emph{not} to be confused with $
\epsilon$ in \eqref{defeps}) is a numerical coefficient, which in the study of~\cite{Dorlis:2024uei} is found to be of order $\mathcal O (10^{-2}$), and~the Hubble parameter $H \approx H_I = constant$, 
satisfying \eqref{HPl}. As~we shall demonstrate below, the~order of magnitude of \eqref{Hbdot} will not be affected by the presence of the $\lambda$-dependent term in \eqref{seaE}, whose contributions to gCS condensate are extremely~suppressed.

To see this, 
we should use perturbation theory in $\dot b$ when evaluating the gCS condensate, following~\cite{Dorlis:2024yqw}.
To this end, formally, we 
 consider weak tensor (GW) perturbations to the FLRW metric, and~work in the transverse--traceless (TT) gauge~\cite{Donoghue:1994dn,tHooft:1974toh}, which implies that only their spatial components, 
 $h_{ij}$, $i,j=1,2,3$, are present in the 
 gauge-fixed effective action. The~respective (3+1)-dimensional line element is of the following form:
\begin{equation}
    ds^2 = - dt^2 + a^2(t) \Big(\delta_{ij} + h_{ij}\Big)\, dx^i \, dx^j\,,
\end{equation}
with $a(t)$ as the scale~factor.

On the helicity ($L,R$) basis, the~metric reads
\begin{equation}
    g_{\mu \nu} = \begin{pmatrix}
        -1 & 0 & 0 & 0 \\
        0 & a^2(t) \left(1 + \frac{h_L (t,z) + h_R (t,z)}{\sqrt{2}} \right) & i a^2(t) \left(\frac{h_L (t,z) - h_R (t,z)}{\sqrt{2}} \right) & 0 \\
        0 & i a^2(t) \left(\frac{h_L (t,z) - h_R (t,z)}{\sqrt{2}} \right) & a^2(t) \left(1 - \frac{h_L (t,z) + h_R (t,z)}{\sqrt{2}} \right) & 0 \\
        0 & 0 & 0 & a^2(t)
    \end{pmatrix}
\end{equation}
The helicity basis $h_L, h_R$ is related to the real linear polarization basis $h_+, h_\times$ as
\begin{equation}
    h_L = \frac{h_+ + i h_\times}{\sqrt{2}}, h_R = \frac{h_+ - i h_\times}{\sqrt{2}}
\end{equation}
We consider the GW perturbations $h_{ij}$ to be small, propagating along the $z$ direction for concreteness. 
We thus obtain the 
following expression for the gCS condensate in~the linearized approximation and in~conformal time (the derivative with respect to which are denoted by primes)~\cite{Alexander:2004us,Lyth:2005jf,Dorlis:2024yqw}:
\begin{equation}
    \langle R_{\mu\nu\rho\sigma} \, \Tilde{R}^{\mu\nu\rho\sigma}\rangle = - \frac{2i}{a^4} \, \left[ \langle\partial^{2}_z h_L\partial_z h_{R}^{\prime}\rangle+ \langle h^{\prime\prime}_L\partial_z h^{\prime}_R\rangle -\langle\partial^{2}_z h_R\partial_z h^{\prime}_L\rangle- \langle h^{\prime\prime}_R\partial_z h^{\prime}_L \rangle\right]\,,
    \label{RCSconformal}
\end{equation}
As discussed in~\cite{Dorlis:2024yqw}, the~quantity inside the square brackets $\left[ \dots \right]$ on the right-hand side is independent of the scale factor $a$ of the Universe; thus, the gravitational CS condensate \eqref{RCSconformal} is inversely proportional to the fourth power of the scale factor $a$. 

To estimate \eqref{RCSconformal}, we first need to determine solutions to the linearized GW equations in~the modified effective action \eqref{seaE}, 
which read:
\begin{align}\label{hLeom}
    &\left[-\left(1 + \lambda \kappa^2 \dot{b}^2 \right) \partial^2_t - \left(3 \frac{\dot{a}}{a} \left(1 + \lambda \kappa^2 \dot{b}^2 \right) + 2 \lambda \kappa^2 \dot{b} \ddot{b} \right) \partial_t + \frac{1}{a^2} \partial_z^2 \right] h_L (t,x) \nonumber \\&= - 2 i \epsilon \kappa^2 \left[\frac{1}{a^2} \left(2 \dot{a} \dot{b} - a \ddot{b} \right) \partial_t \partial_z + \frac{1}{a} \dot{b} \partial^2_t \partial_z - \frac{1}{a^3} \dot{b} \partial^3_z \right] h_L (t,x)\,,
\end{align}
\begin{align}\label{hReom}
    &\left[-\left(1 + \lambda \kappa^2 \dot{b}^2 \right) \partial^2_t - \left(3 \frac{\dot{a}}{a} \left(1 + \lambda \kappa^2 \dot{b}^2 \right) + 2 \lambda \kappa^2 \dot{b} \ddot{b} \right) \partial_t + \frac{1}{a^2} \partial_z^2 \right] h_R (t,x) \nonumber \\ &=  2 i \epsilon \kappa^2 \left[\frac{1}{a^2} \left(2 \dot{a} \dot{b} - a \ddot{b} \right) \partial_t \partial_z + \frac{1}{a} \dot{b} \partial^2_t \partial_z - \frac{1}{a^3} \dot{b} \partial^3_z \right] h_R (t,x)\,.
\end{align}
We observe from \eqref{hLeom} and \eqref{hReom} that the presence of the Chern--Simons coupling causes the equations for left and right waves to be different, a~phenomenon known as cosmological birefringence. On~the other hand, as~expected, the~$\lambda$-dependent corrections \eqref{alternateRicci} do not exhibit that property. Thus, such terms alone (i.e.,~in the absence of the gravitational anomaly CS term, when $\epsilon =0$) yield zero contributions to \eqref{RCSconformal}. Of~course, as~we shall argue below, there are non-trivial but~suppressed $\lambda$-dependent corrections if $\epsilon \ne 0$.

To  estimate these corrections, we should first solve, following~\cite{Dorlis:2024yqw,Dorlis:2024uei}, the above equations in~order to determine the so-called mode functions for the two GW polarizations. Then, upon~canonical quantization of $h_{L,R}$, which now become quantum operators $\widehat h_{L,R}$, we can estimate the vacuum expectation value of \eqref{RCSconformal} with respect to an appropriate quantum vacuum state, which, as~in the $\lambda=0$ case of~\cite{Dorlis:2024yqw,Dorlis:2024uei}, is taken to be the Bunch--Davies vacuum. This procedure will lead to a non-vanishing vacuum approximately constant expectation value for the Pontryagin term \eqref{RCSconformal} during the inflationary period (indicated by the suffix ``$I$'')~\cite{bms,Dorlis:2024yqw,Dorlis:2024uei}:
\begin{equation}\label{gCScond}
   a^4\,  \langle R_{\mu\nu\rho\sigma} \, \Tilde{R}^{\mu\nu\rho\sigma}\rangle_I = {\rm constant} \ne 0\,,
\end{equation}
This, due to the above discussion, is expected to be proportional to $\epsilon$.   

Below, we provide a rather heuristic approach for the estimation of the order of the $\lambda$-dependent-term contribution to the gCS condensate (for a more rigorous approach, see Appendix \ref{sec:confgCS}).
To this end, we first assume an approximately constant $\dot b$ during inflation in~which $H \simeq  H_I = {\rm constant}$, with~the magnitude of $H_I$ given by observations~\cite{Planck:2018vyg}, \eqref{HPl}. In~such a case, terms involving $\ddot b$ 
in the chiral GW equations of motion \eqref{hLeom} and \eqref{hReom} can be ignored. 
By redefining
\begin{align} \label{arescale}
a(t) \quad &\rightarrow \quad {a}^\lambda(t) = a(t)\, \sqrt{1 + \lambda \kappa^2 \dot b^2}\,
\end{align}
the Equations \eqref{hLeom} and \eqref{hReom}, after~straightforward manipulations, can be 
approximated~by
\begin{align}\label{hLeomappr}
    &\Big[- \partial^2_t - 3 \frac{\dot{a}^\lambda}{a^\lambda}  \partial_t + \frac{1}{(a^\lambda)^2} \partial_z^2 \Big] {h}_L (t,x) \nonumber \\& \simeq  - 2 i \, \epsilon \kappa^2 \, \dot b \,\sqrt{1 + \lambda \kappa^2 \dot b^2} \,   \left[ 2\,\frac{\dot{a}^\lambda}{(a^\lambda)^2} \,  \partial_t \partial_z + \frac{1}{a^\lambda} \, \partial^2_t \partial_z - \frac{1 + \lambda \kappa^2 \dot b^2}{(a^\lambda)^3} \, \partial^3_z \right] {h}_L (t,x)\,,
\end{align}
\begin{align}\label{hReomappr}
    &\Big[- \partial^2_t - 3 \frac{\dot{a}^\lambda}{a^\lambda}  \partial_t + \frac{1}{(a^\lambda)^2} \partial_z^2 \Big] {h}_R (t,x) \nonumber \\& \simeq   2 i \, \epsilon \kappa^2 \, \dot b \, \sqrt{1 + \lambda \kappa^2 \dot b^2} \, \left[ 2\,\frac{\dot{a}^\lambda}{(a^\lambda)^2} \,  \partial_t \partial_z + \frac{1}{a^\lambda} \, \partial^2_t \partial_z - \frac{1 + \lambda \kappa^2 \dot b^2}{(a^\lambda)^3} \, \partial^3_z \right] {h}_R (t,x)\,,
\end{align}

As discussed in Appendix \ref{sec:confgCS} (cf. \eqref{c11}), for~the StRVM cosmological model of~\cite{bms,ms1,Dorlis:2024yqw,Dorlis:2024uei}, the~quantity 
$\lambda \kappa^2 \dot b^2 = \mathcal O (10^{-11}) \ll 1$.
On the other hand, upon~taking the Fourier transform of the GW linear polarizations $h_{L,R}$, assuming propagation along the $z$ direction, we may replace $\frac{1}{a}\partial_z$ by $i\frac{k}{a}$, where $k$ is the magnitude of the momentum vector of the GW polarization. 
We then have~\cite{bms,Dorlis:2024yqw} 
$\frac{k}{a} \lesssim
\mu$, where $\mu$ is the 
UV cutoff, identified 
with the string scale~\cite{Green_Schwarz_Witten_2012} $M_s$, in~the context of the StRVM in \eqref{LMs}
. For~phenomenological reasons, specifically in order for the inflation in the StRVM to have a lifetime in the order of 50-60 e-foldings~\cite{Planck:2018vyg}, the~following 
constraint  must be in operation, as~implied by a dynamical system analysis of the StRVM condensate-induced inflation~\cite{Dorlis:2024yqw}:
\begin{align}\label{MPlMs}
 \kappa M_s \simeq 0.2 \,,
 \end{align}
 This is also in agreement with the study in~\cite{bms}.

On account of 
\eqref{LMs}, \eqref{MPlMs}, \eqref{Hbdot}, and \eqref{HPl}, we then obtain that the dimensionless terms 
$\epsilon (\frac{k}{a}) \kappa^2 \dot b \lesssim 8.3 \times 10^{-7}$. Thus, the $\lambda$-dependent terms on the right-hand side of the wave in Equations \eqref{hLeomappr} and \eqref{hReomappr} may be ignored. In~this approximation, the~
resulting equations, which are  expressed in terms of $a^\lambda(t)$ (\eqref{arescale}), become equivalent to the corresponding wave equations of~\cite{Dorlis:2024yqw}. 
Following, then, this analysis, we may estimate the corresponding parity-violating condensate \eqref{RCSconformal}
in our modified gravitational theory \eqref{seaE} as

\begin{equation}
    \langle R_{\mu\nu\rho\sigma} \, \Tilde{R}^{\mu\nu\rho\sigma}\rangle = - \frac{2i}{a^4} (1-\lambda\kappa^2 \dot{b}^2)^2 \left[ \langle\partial^{2}_z h_L\partial_z h_{R}^{\prime}\rangle+ \langle h^{\prime\prime}_L\partial_z h^{\prime}_R\rangle -\langle\partial^{2}_z h_R\partial_z h^{\prime}_L\rangle- \langle h^{\prime\prime}_R\partial_z h^{\prime}_L \rangle\right]\,,
    \label{RCSconformal2}
\end{equation}

where we keep first-order terms in a series expansion in powers of $\lambda$, and~we 
take into account the aforementioned property of the gravitational CS anomaly condensate to depend on the inverse fourth power of the (rescaled) $a^\lambda$ (\eqref{arescale}). From~\eqref{RCSconformal2}, we therefore observe that the $\lambda$-dependent corrections will be suppressed compared to the $\lambda=0$
results of~\cite{bms,Dorlis:2024yqw,Dorlis:2024uei}, and~thus, the corresponding results regarding the inflationary era remain valid to an excellent~approximation. 

We now remark that in the $\lambda=0$ case, which constitutes the zeroth order correction in the $\lambda$ (i.e.,~Regge-slope $\alpha^\prime$ expansion in the string effective action~\cite{Green_Schwarz_Witten_2012}), 
the argumentation of~\cite{Mavromatos:2022xdo}, which was confirmed by the detailed analysis of~\cite{Dorlis:2024uei}, indicates that the gCS condensate~\eqref{gCScond} is proportional to the proper number density of GW sources $\mathcal N_I$ during the RVM inflationary era. In~refs. \cite{Dorlis:2024yqw,Dorlis:2024uei} the Chern--Simons condensate was estimated within a weak quantum gravity path integral formalism about a FLRW~background. 

In view of our result in \eqref{RCSconformal2}, then, in~our modified effective action \eqref{sea2}, the~corresponding gCS anomaly condensate will be given by

\begin{align}\label{CScondl}
    \langle R_{\mu\nu\rho\sigma} \, \Tilde{R}^{\mu\nu\rho\sigma}\rangle_I^{\lambda}  = \langle R_{\mu\nu\rho\sigma} \, \Tilde{R}^{\mu\nu\rho\sigma}\rangle_I^{\lambda=0} (1-2\lambda \kappa^2 \dot{b}^2)  =    
    \mathcal N_I \,(1-2\lambda \kappa^2 \dot{b}^2) \,
\frac{\epsilon}{8\, \pi^2}\, \kappa^4 \, \mu^4 \, \dot{\overline{b}}\, H^3  + \dots\,,
\end{align}
to leading order in the slow-roll parameter 
$\alpha^\prime \,\kappa\, H \,  {\dot {\overline b}} \, \ll \, 1$ of the double-perturbative expansion, in~powers of $\alpha^\prime $ and $\dot b$,
where $H$ is the approximately constant Hubble parameter 
during inflation, satisfying the observational bound \eqref{HPl}.

As already mentioned, the validity of \eqref{qk} is understood. Thus, from~\eqref{CScondl}, one observes that the effects of the presence of the $\lambda$-dependent terms in the modified gravitational action \eqref{seaE}, as~compared to the $\lambda=0$ case of~\cite{bms,ms1,Dorlis:2024yqw}, represent merely a very mild screening (reduction) of the strength of the effect of the GW source terms $\mathcal N_I^\prime \simeq \mathcal N_I (1 - 2\lambda \, \kappa^2\, \dot b^2)$.

The quantity $\mu$ in \eqref{CScondl} is a UV cutoff of the GW modes, and~
the $\dots$ denote corrections implied by the last term of the integrand of \eqref{seaE}, which we have just seen are~subleading. 

The existence of this condensate means that at~a quantum level, the~gravitational Chern--Simons term in the effective action can be expanded about the condensate as%

\begin{equation}\label{expact}
    \int d^4 x \sqrt{-g} b \, R_{\mu\nu\rho\sigma}\, \widetilde R^{\mu\nu\rho\sigma} = \int d^4 x \sqrt{-g} b \, \langle R_{\mu\nu\rho\sigma}\, \widetilde R^{\mu\nu\rho\sigma}\rangle + \int d^4 x \sqrt{-g} :b \, R_{\mu\nu\rho\sigma}\, \widetilde R^{\mu\nu\rho\sigma}:
\end{equation}
where the $\int d^4 x \sqrt{-g} :b \, R_{\mu\nu\rho\sigma}\, \widetilde R^{\mu\nu\rho\sigma}:$ represent quantum fluctuations, for~which:
$\langle: \dots : \rangle = 0$. As~explained in~\cite{Dorlis:2024yqw}, we can then write a ``re-classicalized'' effective action that has the~form

\begin{equation}
    S_E = \int \bigg[ \frac{1}{2 \kappa^2} R - \frac{1}{2} \partial_\mu b \partial^\mu b \, + \, \frac{\epsilon}{8} b \langle R_{\mu\nu\rho\sigma}\, \widetilde R^{\mu\nu\rho\sigma}\rangle + \frac{\epsilon}{8} :b \, R_{\mu\nu\rho\sigma}\, \widetilde R^{\mu\nu\rho\sigma}: - \frac{\lambda}{2} \partial_\mu b \partial_\nu b R^{\mu \nu} \bigg]\sqrt{-g} \, d^4 x.
\end{equation}
The key here is the addition of a linear-axion (monodromy) potential term $\sim b \langle R_{\mu\nu\rho\sigma}\, \widetilde R^{\mu\nu\rho\sigma}\rangle$, which, as~we will see, drives inflation. The~metric equations of motion become
\begin{equation}
    \frac{1}{2 \kappa^2} G_{\mu \nu} + \frac{\epsilon}{16} \langle R_{\mu\nu\rho\sigma}\, \widetilde R^{\mu\nu\rho\sigma}\rangle g_{\mu \nu}= \frac{\epsilon}{2} \mathcal{C}_{\mu \nu} + \frac{\lambda}{2} \Theta_{\mu \nu} + \frac{1}{2} {T_{\rm kin}^b}_{\mu \nu},
\end{equation}
while the scalar equation of motion reads
\begin{equation}
\nabla_\mu\left(\partial^\mu b + \lambda R^{\mu\nu}\partial_\nu b + \frac{\epsilon}{8} \mathcal{K}^\mu\right) = - \frac{\epsilon}{8} \,  \langle R_{\mu\nu\rho\sigma}\, \widetilde R^{\mu\nu\rho\sigma}\rangle
\end{equation}

Upon expanding the effective action \eqref{expact} about the condensate, the following is implied in general:
\begin{align}\label{expK0}
\mathcal K^0 = \langle \mathcal K^0 \rangle + : \mathcal K^0 : \,,
\end{align}
where, as~already mentioned, the~$: \dots :$ denote quantum fluctuations, characterized by a zero vacuum expectation value, $\langle : \dots : \rangle =0$\footnote{In the case of the gravitational CS term, the~
quantum fluctuations do not contribute to the equations of motion of graviton in the FLRW background.}. 

A constant condensate $\mathcal K^0$ (which spontaneously violates Lorentz symmetry~\cite{bms} if~we view the condensate as a vacuum expectation value of the topological current $\langle \mathcal K^0 \rangle = \rm constant \ne 0$) is  consistent with inflation, characterized by $ H = \dot a / a = {\rm constant}$ and $\ddot a / a = {\rm constant} $, which, on~account of \eqref{bdotK0}, would also imply $\dot{b} = {\rm constant}$, as~in the case of~\cite{bms}. 
 The reader should notice that \eqref{Hbdot} implies a $H^4$ scaling of the Chern--Simons condensate \eqref{CScondl} during~inflation.

Upon concentrating on the condensate (which behaves classically) and~ignoring the fluctuations, we obtain the following
from \eqref{scalar_eom} and \eqref{CScurrent}:
{\small\begin{align}
    \frac{d}{dt} \langle \mathcal 	K^0 (t) \rangle + 3H\, \langle \mathcal K^0 \rangle &= \langle R_{\mu\nu\rho\sigma}\, \widetilde R^{\mu\nu\rho\sigma} \rangle^\lambda_I  \, \nonumber \\ & \stackrel{\eqref{bdotK0},\eqref{CScondl}}{=}\,  
     \mathcal N_I \, (1-2\lambda \kappa^2 \dot{b}^2) \,
    \frac{\epsilon^2}{64\, \pi^2}\, \kappa^4 \, \mu^4 
    \frac{H^3}{\Big(1 + 3\, \lambda\, H^2 \Big)} \, \langle {\mathcal K}^0 \rangle \nonumber \\
    & \simeq  \mathcal N_I \, (1-2\lambda \kappa^2 \dot{b}^2) \, (1 - 3 \lambda H^2)\, 
    \, \frac{\epsilon^2}{64\, \pi^2}\, \kappa^4 \, \mu^4 \, H^3 \,  \nonumber \\ & \stackrel{\eqref{Hbdot}}{\simeq} \, \mathcal N_I \, (1 - 3.04 \, \lambda \, H^2)\, \frac{\epsilon^2}{64\, \pi^2}\,\kappa^4 \, \mu^4 \, H^3 \,,
    \label{diff_eq_K0} 
\end{align}}%
where we expand to linear order in $\lambda$. Equation~\eqref{diff_eq_K0} can be solved to yield at a time $t_{\rm begin} < t < t_{\rm end}$ during inflation, where $t_{\rm begin}$ ($t_{\rm end}$) denotes the onset (end) of the inflationary era:

\begin{align}\label{Ksol}
\langle \mathcal K^0(t) \rangle = \langle \mathcal K^0(t_{\rm begin}) \rangle  \, \exp\Big(-3 H \, (t_{\rm end} - t_{\rm begin}) \Big[1 - \mathcal N_I \, (1 - 3.04 \, \lambda \, H^2 )\, 
    \frac{\epsilon^2}{192\, \pi^2}\, \kappa^4 \, \mu^4 
    H^2 \,\Big]\Big)
\end{align}
The exponent vanishes, as~is expected for a condensate, for~{\small\begin{align}\label{condition}
 1 &= \mathcal N_I \, (1 - 3.04\, \lambda \, H^2) \, \frac{\epsilon^2}{192\, \pi^2} (\kappa \mu)^4 \, H^2\,  \stackrel{\eqref{defeps}}{\simeq}  18.55 \times 10^{-7} \, \mathcal N_I\, (1 - 3.04\, \lambda \, H^2)  \, \kappa^2 H^2  \, \nonumber \\ 
 &\Rightarrow \mathcal N_I = 5.4 \times 10^5 \, \Big(\kappa^{-2}\, H^{-2} + 3.04\, \frac{\lambda}{\kappa^2} \Big)
 \stackrel{\eqref{HPl}}{\Rightarrow} \,  \mathcal N_I \gtrsim 5.4 \times 10^{15} \,,
\end{align}}%
given that $3.04 \, \frac{\lambda}{\kappa^2} \ll \kappa^{-2}\, H^{-2} \sim 10^{10}$ due to \eqref{defeps} and \eqref{HPl}. 

This provides a justification for the approximate constancy of $\mathcal K^0$ employed in our argumentation above. 
The initial value then of $\mathcal K^0(t_{\rm begin})$ at the onset of the RVM inflation in \eqref{Ksol} 
is determined from 
\eqref{bdotK0} by using the parametrization \eqref{Hbdot} on the left-hand side, along with the 
dynamical system analysis estimate of the order of the parameter $\varepsilon = \mathcal O(10^{-2})$~\cite{Dorlis:2024uei}.

We thus observe that the higher-order $\lambda$-dependent corrections in the effective action~\eqref{seaE} do not affect at all the lower bound of $\mathcal N_I$, as~compared to the $\lambda=0$ case, due to the extreme weakness of these terms.
Equation~\eqref{condition} is compatible with the corresponding result in~\cite{Dorlis:2024yqw} based on a dynamical system analysis, where one finds
\begin{align}\label{sources}
 \frac{\mathcal N_I}{\mathcal N_{\rm stiff}} \sim 7 \times 10^{16}\,,
 \end{align}
 which ensures continuity of the value of the Chern--Simons anomaly condensate during the transition from the stiff to inflationary eras ($\mathcal N_{\rm stiff}$ denotes the assumed numerical density of sources during the stiff era that precedes the inflationary epoch in the StRVM~\cite{bms,ms1,Dorlis:2024uei}).  
The equality in \eqref{condition} is satisfied in order of magnitude by $\mathcal N_{\rm stiff} = \mathcal O(10^{-1})$. So, the conclusions of the previous approach of~\cite{bms,Dorlis:2024yqw} remain qualitatively and quantitatively correct, upon~the validity of \eqref{HPl}, which is supported by the Planck data phenomenology of inflation~\cite{Planck:2018vyg}. 

A remark we would like to make at this point concerns the consistency of the condensate inflation with classical field theory considerations. Indeed, despite the fact that the condensate is the result of quantum GW primordial excitations, which condense,  its value remains nonetheless consistent with the Euler--Lagrange (classical) equation of motion for the axion-like field $b$ after~the formation of the condensate in~an FLRW background:

\begin{align}\label{bflkrw}
\ddot b + 3 H \dot b = \frac{\epsilon}{8} \, 
 \langle R_{\mu\nu\rho\sigma} \, \Tilde{R}^{\mu\nu\rho\sigma}\rangle_I^{\lambda}  \stackrel{\eqref{diff_eq_K0}}{\simeq}    
    \mathcal N_I \,(1-2\lambda \kappa^2 \dot{b}^2) \,
\frac{\epsilon^2}{64\, \pi^2}\, \kappa^4 \, \mu^4 \, \dot{b}\, H^3 \simeq  \mathcal N_I \,
\frac{\epsilon^2}{64\, \pi^2}\, \kappa^4 \, \mu^4  \, \dot{b}\, H^3 \,,
\end{align}
where in the last line, we ignored the subleading $\lambda$-dependent contributions. To~check the validity of this equation, during~the condensate-induced inflation, we rewrite it for~an approximate constant, i.e.,~$\dot b =0$, as follows:
\begin{align}\label{bflrw}
 3 \simeq  \mathcal N_I \,
\frac{\epsilon^2}{64\, \pi^2}\, \kappa^4 \, \mu^4 \, H^2 \,. 
\end{align}
Using \eqref{HPl}, we obtain the following from \eqref{bflrw}:
\begin{align}
    \mathcal N_I \gtrsim 1.4 \times 10^{15}\,,
\end{align}
which is remarkably consistent with 
the bound \eqref{condition}, if~one takes into account the theoretical uncertainties involved, due to the dominant role of GW modes near the UV cutoff $M_s$ in the formation of the gCS anomaly condensate \eqref{gCScond}. 

The above result indicates that higher-derivative and higher-than-quadratic curvature terms in the string effective action, which were naively expected to play an important role for modes with momenta near the UV cutoff, will not affect (in order of magnitude) the analysis of~\cite{bms,ms1,Dorlis:2024yqw} for a primordial GW condensate-induced inflation, based on lowest-non-trivial-order $\mathcal O(\alpha^\prime)$ string effective actions with gravitational anomalies. The~consistency of the condensate approach from many rather distinct viewpoints, ranging from dynamical system analysis to~weak quantum gravity effective field theory, offers a non-trivial support to~this. 

Finally, before~concluding this section, we remark that as~shown in~\cite{Dorlis:2024uei}, one can achieve an extremely good agreement in the phenomenology of the gCS condensate inflationary scenario described above with the standard scale invariance violating small slow-roll inflationary parameters, $n_s, \varepsilon$, and $\eta$, as~defined in the standard inflationary phenomenology~\cite{Planck:2018vyg}
$\varepsilon_1 = \frac{1}{2\, \kappa^{2}} \Big(\frac{V^\prime}{V}\Big)^2$, with~the prime denoting a derivative with respect to the field $b$, $\eta = \frac{1}{\kappa^{2}}\,  \frac{V^{\prime\prime}}{V}$, and~the running spectral index 
$n_s = 1 - 6\, \varepsilon_1 + 2\, \eta$. In~addition, the~ratio of the scalar to tensor perturbations $r= 16\, \varepsilon$.
Indeed, by~considering instanton effects in the non-Abelian gauge sector that may characterize the model, one obtains an effective, instanton-induced periodic modulation of the axion $b$ potential as follows:
\begin{align}\label{axionpot}
V(b)_{\rm eff} \ni \Lambda_1^4\, \cos\Big(2\, \pi^2\, \epsilon \, b(x) \Big) \equiv \Lambda_1^4\, \cos\Big(\frac{b}{f_b}\Big)\,, \end{align}
where $\Lambda_1$ is the instanton energy scale, and~$\epsilon$ is defined in \eqref{defeps}. 
The quantity $f_b$ is the axion $b$ coupling, defined through the gauge sector of the Prontryagin  anomaly term in the action~\cite{EGUCHI1980213}, upon~the inclusion of the gauge sector, as follows~\cite{Dorlis:2024uei}:
\begin{align}\label{bff} \mathcal S_{\rm anom} \ni \frac{1}{16\pi^2\, f_b} \int d^4x \sqrt{-g}\, b(x) \, 
{\mathbf F}_{\mu\nu}\, \widetilde{{\mathbf F}}^{\mu\nu}\,,
\end{align}
where ${\mathbf F}_{\mu\nu}$ is the non-Abelian-gauge-group field strength and 
$\frac{1}{16\pi^2} \int d^4x \sqrt{-g}\,  
{\mathbf F}_{\mu\nu}\, \widetilde{{\mathbf F}}^{\mu\nu} =n$, $n \in \mathbb Z$, the~Pontryagin index~\cite{EGUCHI1980213}. By~tuning appropriately the scale $\Lambda_1$ to certain natural values and~assuming $\frac{\kappa}{\sqrt{\alpha^\prime}} = {\mathcal O}(0.1) $, as~dictated by the inflationary phenomenology of the model of~\cite{Dorlis:2024yqw} (i.e.,~the requirement that the duration of inflation be in the region of $60-70$ e-foldings), we obtain~\cite{Dorlis:2024uei}, a per-mil agreement
on the values of the slow-roll parameters of the model with those measured by the Planck collaboration~\cite{Planck:2018vyg}.

\section{Conclusions}\label{sec:concl}

In this work, we revisited the string-inspired running vacuum model (StRVM) of cosmology proposed in~\cite{bms,ms1,ms2,Dorlis:2024yqw,Dorlis:2024uei} by~considering the most general low-energy effective gravitational action quartic in spacetime derivatives of graviton and antisymmetric tensor fields (assuming a stabilized dilaton to a constant value, which sets the scale for the string coupling, and~hence the relevant phenomenology). 
In (3+1)-dimensional spacetimes, we have shown that it is possible to 
 simultaneously impose unitarity (i.e.,~a Gauss--Bonnet-quadratic curvature scheme, which ensures absence of graviton ghosts) and~torsional interpretation of the field strength of the antisymmetric tensor field, which behaves as a totally antisymmetric component of spacetime~torsion.

By carefully studying  the potential field redefinitions
leading to the above features,
we managed to arrive at an appropriate basis of structures that contains only one extra term of the fourth-order in derivatives, as~compared to the effective action of the StRVM, of~the form 
$$  \mathcal S_\lambda^{\rm extra} = \lambda \, \int d^4x \, \sqrt{-g}\, \partial_\mu b \, \partial_\nu b \, R^{\mu\nu}\,, \qquad \lambda \propto \alpha^\prime \,,$$where $R_{\mu\nu}$ denotes the torsion-free Ricci tensor of the spacetime geometry, and~$b$ is the massless axion-like field,  which, as~discussed in Section \eqref{sec:Haxion} (cf. \eqref{friedan}), is expected from the seminal work of~\cite{Friedan:1980jf,Friedan:1980jm}. The~axion field $b(x)$  is dual (in the (3+1)-dimensional spacetime after string compactification) to the field strength (and totally antisymmetric torsion) $\mathcal H_{\mu\nu\rho}$ in the sense of \eqref{dual3} when order $\alpha^\prime$ terms in the effective action are taken into account. Nonetheless, for~Einstein spaces (as is the inflationary spacetime of interest here), this modified duality relation maintains the essential structure of \eqref{dual}, but~now includes a global proportionality scalar factor, dependent on the scalar spacetime curvature (cf. \eqref{Adef}, \eqref{einsteinH}), as follows: 
$$ 
\mathcal H^{\rm Einstein}_{\mu\nu\rho} = 3 \Big(1 - \frac{0.015\, \alpha^\prime}{2}\, R\Big)\, \eta_{\mu\nu\rho\sigma}\, \partial^\sigma b
\,.$$

We explained in the main text the way the  above $\lambda$-dependent term $\mathcal S_\lambda^{\rm extra}$  arises after~path-integrating the torsion $\mathcal{H}_{\mu\nu\rho}$ field in  the partition function with respect to the generalized four-derivative effective action \eqref{string_effaction}. The~axion field is introduced in the path integral as a Bianchi identity constraint implementing a Lagrange multiplier field, which is originally non-dynamical. 
The dynamics of $b(x)$ arise after the Gaussian path integration over the $\mathcal H$-torsion.

The parameter $\lambda \propto \alpha^\prime = M_s^{-2}$, and~this implies that the contribution of the parity-even term in the parity-odd gravitational anomaly condensate is negligible for~the values of string scale $M_s$ required for the correct phenomenology of the StRVM inflation~\cite{Dorlis:2024yqw}. This leads to the conclusion that the analysis of the StRVM, which ignored such $\lambda$-dependent contributions, is complete in~view of the extreme suppression of these~terms.

We  also discussed a remarkable consistency of the anomaly condensate-induced inflation in the context of the StRVM, which stems perhaps from the exactness of the anomaly term per se. Indeed, we have shown how the value of the gravitational anomaly condensate, as~estimated by weak-chiral-GW quantum tensor perturbation effective field theory, is consistent with the satisfaction of the classical Euler--Lagrange equations of motion of the axion $b(x)$ field, thereby pointing towards a ``classical'' nature of the anomaly condensate. 
As discussed in~\cite{bms}, this is also consistent with a spontaneous violation of Lorentz symmetry due to the formation of a vacuum epxectation value of the temporal component of the 
topological current corresponding to the Chern--Simons gravitational anomaly
term as~a consequence 
of the GW-induced anomaly~condensation.

\appendix

    \section{Decomposition of the Generalized (Contorted) Curvature Tensors and Useful Identities}\label{sec:identities}

 Let us assume a torsionful connection of the following form:
\begin{equation}
    \bar{\Gamma}^\lambda_{\mu\nu} = \Gamma^\lambda_{\mu\nu} + \frac{\kappa}{\sqrt{3}} \mathcal{H}^\lambda{}_{\mu \nu},
\end{equation}
In this case, $\frac{\kappa}{\sqrt{3}} \mathcal{H}^\lambda{}_{\mu \nu}$ plays the role of the contorsion tensor $K^\lambda{}_{\mu \nu}$. The~torsion tensor $T^\lambda{}_{\mu \nu}$ is given in terms of the contorsion tensor as
\begin{equation}
    {T^a}_{bc} = 2 {K^a}_{[bc]} = ({K^a}_{bc} - {K^a}_{cb}).
\end{equation}
Therefore, in~our case, for~a totally antisymmetric contorsion, the~torsion tensor is given as
\begin{equation}
    T_{\mu \nu \lambda} = \frac{2 \kappa}{\sqrt{3}} \mathcal{H}_{\mu \nu \lambda}.
\end{equation}
We can then use the above to ``decompose'' the generalized curvature (torsionful Riemann) tensor into the usual (Levi--Civita connection) Riemann tensor and the totally antisymmetric torsion $\mathcal{H}$ using the following relations:
\begin{equation}
    \bar{R}_{\mu \nu \lambda \rho} = R_{\mu \nu \lambda \rho} + \frac{\kappa}{\sqrt{3}}\left( \nabla_\lambda \mathcal{H}_{\mu \nu \rho} - \nabla_\rho \mathcal{H}_{\mu \lambda \nu }\right) + \frac{\kappa^2}{3}\left( \mathcal{H}_{\mu \lambda \sigma} \mathcal{H}^\sigma{}_{\rho \nu} - \mathcal{H}_{\mu \rho \sigma} \mathcal{H}^\sigma{}_{\lambda \nu}\right)
\end{equation}
\begin{equation}
    \bar{R}_{\mu \nu} = R_{\mu \nu} - \frac{\kappa}{\sqrt{3}} \nabla_\lambda \mathcal{H}_{\mu \nu}{}^\lambda + \frac{\kappa^2}{3} \mathcal{H}_{\mu \lambda}{}^\sigma \mathcal{H}^\lambda{}_{\nu \sigma}
\end{equation}
\begin{equation}
    \bar{R} = R - \frac{\kappa^2}{3} \mathcal{H}^{\mu \nu \lambda} \mathcal{H}_{\mu \nu \lambda}
\end{equation}
For a totally antisymmetric torsion tensor, the~first Bianchi identity for the Levi--Civita Riemann tensor, $R_{\mu[\nu \lambda \rho]}=0$, results in
\begin{equation}
    \mathcal{H}_{\kappa}{}^{\mu \nu} \mathcal{H}^{\kappa \lambda \rho} R_{\mu \lambda \nu \rho}  = \frac{1}{2} \mathcal{H}_{\kappa}{}^{\mu \nu} \mathcal{H}^{\kappa \lambda \rho} R_{\mu \nu \lambda \rho} 
\end{equation}
The second Bianchi identity, $\nabla_{[\sigma|}R_{\mu \nu | \lambda \rho]}=0$, results in
\begin{equation}
    \int R_{\mu \nu \lambda \rho} \nabla^{\rho} \mathcal{H}^{\mu \nu \lambda} \sqrt{-g} \, d^4 x = \int \nabla^\rho(R_{\mu \nu \lambda \rho} \mathcal{H}^{\mu \nu \lambda}) \sqrt{-g} \, d^4 x
\end{equation}
Through partial integrations and commuting covariant derivatives, we also get
\begin{equation}
    \begin{split}
        \int \nabla_{\lambda} \mathcal{H}_{\mu \nu \rho} \nabla^{\rho} \mathcal{H}^{\mu \nu \lambda} \sqrt{-g} \, d^4 x &= \int \nabla_\lambda ( \mathcal{H}_{\mu \nu \rho} \nabla^\rho \mathcal{H}^{\mu \nu \lambda} - \mathcal{H}^{\mu \nu \lambda \rho} \nabla^d \mathcal{H}_{\mu \nu \rho}) \sqrt{-g} \, d^4 x \\
        &+ \int \nabla_\lambda \mathcal{H}^{\mu \nu \lambda} \nabla^\rho \mathcal{H}_{\mu \nu \rho} \sqrt{-g} \, d^4 x - \int \mathcal{H}_\mu{}^{\lambda \rho} \mathcal{H}_{\nu \lambda \rho} R^{\mu \nu} \sqrt{-g} \, d^4 x \\
        &+ \int \mathcal{H}_{\kappa}{}^{\mu \nu} \mathcal{H}^{\kappa \lambda \rho} R_{\mu \nu \lambda \rho} \sqrt{-g} \, d^4 x 
    \end{split}
\end{equation}
The Bianchi identity for the Kalb--Ramond field strength, $\nabla_{[\rho} \mathcal{H}_{\mu \nu \lambda]} = 0$, results in
\begin{equation}
    \mathcal{H}_{\mu}{}^{\rho \sigma} \mathcal{H}^{\mu \nu \lambda} \nabla_{\sigma} \mathcal{H}_{\nu \lambda \rho} = 0
\end{equation}
and
\begin{equation}
    \begin{split}
        \int \nabla_{\rho} \mathcal{H}_{\mu \nu \lambda} \nabla^{\rho} \mathcal{H}^{\mu \nu \lambda} \sqrt{-g} \, d^4 x  &= \int \nabla_\rho (\mathcal{H}_{\mu \nu \lambda} \nabla^\rho \mathcal{H}^{\mu \nu \lambda} - 3 \mathcal{H}^{\mu \rho \nu} \nabla^\lambda \mathcal{H}_{\mu \nu \lambda}) \sqrt{-g} \, d^4 x \\
        &+ 3 \int \nabla_\lambda \mathcal{H}^{\mu \nu \lambda} \nabla^\rho \mathcal{H}_{\mu \nu \rho} \sqrt{-g} \, d^4 x - 3 \int \mathcal{H}_{\mu}{}^{\lambda \rho} \mathcal{H}_{\nu \lambda \rho} R^{\mu \nu} \sqrt{-g} \, d^4 x \\
        &+ 3 \int \mathcal{H}_{\kappa}{}^{\mu \nu} \mathcal{H}^{\kappa \lambda \rho} R_{\mu \nu \lambda \rho} \sqrt{-g} \, d^4 x \,.
    \end{split}
\end{equation}

 \section{Field Redefinitions Directly in D = 4 Field Theory Effective Actions Beyond~String Theory}\label{sec:beyondstrng}

In this Appendix, for~completeness, we shall give the result of the field redefinitions~\eqref{fred} and \eqref{fred2} directly to a $D=4$ field theory action, which formally has  the same form as the $\mathcal O(\alpha^\prime)$ string action \eqref{act0} (\eqref{act1}, \eqref{act2}) but, here, represents a four-dimensional four-derivative field theory action, independent of strings, for~which the standard S-matrix equivalence theorems~\cite{Kamefuchi:1961sb,Salam:1970fso,Kallosh:1972ap,Bergere:1975tr,Georgi:1991ch,Weinberg:1995mt} apply. In~such a case, the parameter
\begin{align}\label{alphaM}
    \sqrt{\alpha^\prime}=\frac{1}{\mathcal M}\,,
    \end{align}
    may be identified with the inverse of an energy scale below which the effective field theory is valid, i.e.,~an UV cutoff, not related to the string case. 
Below, we shall outline the relevant differences between the two formalisms and~show that in the context of a cosmological model, like the StRVM, the~basic conclusions are not affected in order of~magnitude. 

Should one perform the field redefinitions directly in $D=4$, there are different identities for the torsion terms that can be used, as~compared to the $D > 4$ case studied in Section~\ref{sec:redef}. 
In this case, the~expression \eqref{unitary_string_action_general_D=4_old} becomes

\begin{equation}
    \begin{split}
        (\delta S_0 + S_1)^{D=4}_{\text{Unitary}} &= \int \bigg\{ \frac{1}{16 \kappa^2} \left( R_{\mu \nu \lambda \rho} R^{\mu \nu \lambda \rho} - 4 R_{\mu \nu} R^{\mu \nu} + R^2 \right) + \\
        &\phantom{=} - \frac{1}{8} \bigg[ \left( 4 B_3 - \frac{2}{3 \sqrt{3}}\right) \mathcal{H}_{\mu \lambda \rho} \mathcal{H}_{\nu}{}^{\lambda \rho} R^{\mu \nu} + \left( \frac{1}{9\sqrt{3}} - 2 B_3 - 4 B_4 \right) \mathcal{H}_{\mu \nu \lambda} \mathcal{H}^{\mu \nu \lambda} R \bigg] + \\
        &\phantom{=}+ \frac{\kappa^2}{24} \bigg[ \left(-1-\frac{2}{9 \sqrt{3}} + 12 B_3 + 24 B_4 \right) \mathcal{H}^{\mu \nu \lambda} \mathcal{H}_{\mu}{}^{\rho \sigma} \mathcal{H}_{\nu \rho}{}^\kappa \mathcal{H}_{\lambda \sigma \kappa} \bigg] \bigg\}  \sqrt{-g} \, d^4 x,
    \end{split}
    \label{unitary_string_action_general_D=4_old}
\end{equation}
and matching with \eqref{unitary_action_generalized_curvature_D=4} yields
\begin{align}\label{ABold}
    B_4 &= - \frac{1}{2} B_3 + \frac{1}{24} + \frac{1}{108 \sqrt{3}},\\
    B_3 &= + \frac{1}{8} + \frac{1}{9 \sqrt{3}}, \\
    A_4 &= + \frac{11}{128} - \frac{1}{96 \sqrt{3}}, \\
    A_5 &= - \frac{3}{128} + \frac{1}{96 \sqrt{3}}\,.
\end{align}
instead of \eqref{AB}. 

 The resulting unitary effective string action where $\mathcal H_{\mu\nu\rho}$ also plays the role of torsion~reads

\begin{equation}
    \begin{split}
        (\delta S_0 + S_1)^{D=4}_{\text{Unitary, Torsion}} &= \int \bigg\{ \frac{1}{16 \kappa^2} \left( R_{\mu \nu \lambda \rho} R^{\mu \nu \lambda \rho} - 4 R_{\mu \nu} R^{\mu \nu} + R^2 \right) + \\
        &\phantom{=} - \frac{1}{8} \left(\frac{1}{2} - \frac{2}{9 \sqrt{3}} \right) \bigg[ \mathcal{H}_{\mu \lambda \rho} \mathcal{H}_{\nu}{}^{\lambda \rho} R^{\mu \nu} -\frac{1}{3} \mathcal{H}_{\mu \nu \lambda} \mathcal{H}^{\mu \nu \lambda} R \bigg] \bigg\}  \sqrt{-g} \, d^4 x\,.
    \end{split}
\end{equation}
Since in $D=4$ the Gauss--Bonnet quadratic curvature combination is a total derivative, which is thus ignored, the~full effective action reads
\begin{align}\label{string_effaction_old}
    S = &\int \bigg[\frac{1}{2\kappa^2} R - \frac{1}{6} \mathcal{H}_{\mu \nu \lambda} \mathcal{H}^{\mu \nu \lambda} \nonumber \\
    &+\alpha^\prime \left(\frac{1}{36 \sqrt{3}} - \frac{1}{16} \right) \left( \mathcal{H}_{\mu \lambda \rho} \mathcal{H}_{\nu}{}^{\lambda \rho} R^{\mu \nu} -\frac{1}{3} \mathcal{H}_{\mu \nu \lambda} \mathcal{H}^{\mu \nu \lambda} R\right) \bigg] \sqrt{-g} \, d^4 x\,.
\end{align}
The reader can compare directly \eqref{string_effaction_old} with \eqref{string_effaction}, and~the difference arises in both the~magnitude and signature of the~coefficient of the 
$\mathcal O(\alpha^\prime)$ terms. 

The resulting effective action (in terms of the axion fields that arise after the $\mathcal H$-torsion path integration) maintains its generic form \eqref{seaE}, but~now 
with
\begin{align}\label{defeps2}
\lambda &= \alpha^\prime \left(\frac{1}{8} - \frac{1}{18 \sqrt{3}} \right) = 0.093 \, \alpha^\prime  \, > \, 0 \,, \nonumber\\
    \frac{1}{2q^2} &= \frac{1}{2\kappa^2} \left[ 1 - \frac{\kappa^2}{\alpha^\prime }\frac{\lambda}{\alpha^\prime} \right] = \frac{1}{2\kappa^2} \Big(1 - 0.093 \frac{\kappa^2}{\alpha^2} \Big) \, ,\\
    \epsilon & = \frac{\alpha^\prime}{\kappa} \frac{\sqrt{2}}{12} \nonumber \,,
\end{align}
where now $\alpha^\prime$ is given by \eqref{alphaM}, related to the UV cutoff scale of the effective theory, which is not the string scale. As~this is a gravitational theory, it is natural to identify $\mathcal M$ with the effective Planck scale $q$ \eqref{defeps2}. Due to the positive nature of the parameter $\lambda$ in this case, we observe that the effective Planck energy scale $q^{-1}$ is smaller than the bare one, $\kappa^{-1}$. In~fact, to maintain unitarity of the theory ($q^2 > 0$), we need to impose the following constraint:
\begin{align}\label{constralphaM}
  10.76 \, \kappa^{-2} > \mathcal M^2 \,    
\end{align}
This constraint is naturally satisfied if we impose the transplanckian conjecture, according to which no energy scale in our effective field theory should exceed the Planck scale.
The opposite happens in Section~\ref{sec:Haxion}, \eqref{defeps}, where, because~of the negative nature of the respective parameter $\lambda$, the~effective Planck energy scale $q$ (which is always positive for any $\lambda < 0$) is larger than the bare $\kappa^{-1}$. 

In the context of the D=4 dimensional StRVM, with~the improved effective action \eqref{seaE}, the~value of $0 < \lambda = 0.093 \frac{1}{\mathcal M^2}$ is such that with~the natural identification $\mathcal M = M_{\rm Pl} = q^{-1}$, the~$\lambda$-dependent term in \eqref{seaE} still makes subleading contributions to the gCS condensate in the context of the StRVM, for~which \eqref{Hbdot} and \eqref{HPl} are~valid.
 
 \section{Parity-Odd Quadratic Curvature Invariants in 4D}\label{sec:parityandR2}

In the context of our study of quadratic generalized curvature invariants, it is worth making a short comment about parity-odd quadratic invariants. These are invariants that involve two generalized curvature tensors and the Levi--Civita tensor. Thus, their existence, number, and explicit form are strongly dependent on the number of spacetime dimensions.
It is trivial to see, for~example, that such invariants do not exist in any odd number of dimensions since the number of indices in those is odd and a scalar cannot be formed. In~an even number of dimensions, such invariants exist up until $D=8$ because~this is the maximum number of free indices that two generalized curvature tensors may have. Therefore, the~absence of such invariants in $10D$ heterotic string action should not be a surprise, since they simply do not exist. In~$D=4$, the~number of parity-odd invariants is four~\cite{Baekler:2011jt}:
\begin{align}
    \bar{G}^-_7 &= \bar{R} \eta_{\mu \nu \lambda \rho} \bar{R}^{\mu \nu \lambda \rho} = \bar{R} \Tilde{\bar{R}},\\
    \bar{G}^-_8 &= \eta^{\mu \nu \kappa \sigma} \bar{R}_{\mu \nu \lambda \rho} \bar{R}_{\kappa \sigma}{}^{\lambda \rho} = \bar{R}_{\mu \nu \lambda \rho} \Tilde{\bar{R}}^{\mu \nu \lambda \rho},\\
    \bar{G}^-_9 &= \eta^{\lambda \rho \kappa \sigma} \bar{R}_{\mu \nu \lambda \rho} \bar{R}^{\mu \nu \kappa \sigma},\\
    \bar{G}^-_{10} &= \eta^{\lambda \rho \kappa \sigma} \bar{R}_{\mu \nu \lambda \rho} \bar{R}_{\kappa \sigma}{}^{\mu \nu} = \bar{R}_{\mu \nu \lambda \rho} \Tilde{\bar{R}}^{\lambda \rho \mu \nu},
\end{align}
where we assume a general torsion. Note that $\bar{G}^-_8$ is the topological Pontryagin term, a~total derivative. If~we specify our torsion to be the totally antisymmetric tensor $\mathcal{H}_{\mu \nu \lambda}$, which also satisfies the Bianchi identity $\nabla_{[\rho} \mathcal{H}_{\mu \nu \lambda]} = 0$, these four parity-odd quadratic invariants are evaluated to be modulo total derivative terms as follows:
\begin{align}
    \bar{G}^-_7 &= 0,\\
    \bar{G}^-_8 &= R_{\mu \nu \lambda \rho} \Tilde{R}^{\mu \nu \lambda \rho}, \\
    \bar{G}^-_9 &= R_{\mu \nu \lambda \rho} \Tilde{R}^{\mu \nu \lambda \rho},\\
    \bar{G}^-_{10} &= R_{\mu \nu \lambda \rho} \Tilde{R}^{\mu \nu \lambda \rho}.
\end{align}
Therefore, we find that all of these invariants either collapse to the torsionless Pontryagin term or vanish, which shows that there are no terms that the effective string action (at order $\mathcal{O}(\alpha^\prime)$) does not account for due to its original formulation being in higher~dimensions.

\section{Conformal Time Analysis of \boldmath{$\lambda$}-Dependent Corrections to the gCS Condensate}\label{sec:confgCS}

Going into conformal time, the~wave equations for the chiral GW modes \eqref{hLeomappr} and \eqref{hReomappr} become

\begin{equation}
    \left(1+\lambda\kappa^2 \left(\frac{b'}{a}\right)^2\right)h_L'' + \left(2 \frac{a'}{a} + 2 \lambda \kappa^2 \frac{b'}{a} \frac{b''}{a}\right) h_L' - \partial_z^2 h_L = \frac{2 i \epsilon \kappa^2}{a^2} \partial_z \left( b'' h_L' + b' h_L'' - b' \partial_z^2 h_L\right)
\end{equation}
\begin{equation}
    \left(1+\lambda\kappa^2 \left(\frac{b'}{a}\right)^2\right)h_R'' + \left(2 \frac{a'}{a} + 2 \lambda \kappa^2 \frac{b'}{a} \frac{b''}{a}\right) h_R' - \partial_z^2 h_R = -\frac{2 i \epsilon \kappa^2}{a^2} \partial_z \left( b'' h_R' + b' h_R'' - b' \partial_z^2 h_R\right)
\end{equation}
Going into Fourier modes:
\begin{equation}
    h_{L,R} (\eta, \Vec{x}) = \int \frac{d^3 k}{(2\pi)^{\frac{3}{2}}} e^{i \Vec{k} \cdot \Vec{x}} h_{L,R,\Vec{k}} (\eta),
\end{equation}
and substituting, we get
\begin{align}
    &\left(1+\lambda\kappa^2 \left(\frac{b'}{a}\right)^2\right)h_{L,\Vec{k}}'' + \left(2 \frac{a'}{a} + 2 \lambda \kappa^2 \frac{b'}{a} \frac{b''}{a}\right) h_{L,\Vec{k}}' + k^2 h_{L,\Vec{k}} \nonumber \\ &= - \frac{2 k \epsilon \kappa^2}{a^2} l_{\Vec{k}} \left( b'' h_{L,\Vec{k}}' + b' h_{L,\Vec{k}}'' + k^2 b' h_{L,\Vec{k}}\right)
\end{align}
\begin{align}
    &\left(1+\lambda\kappa^2 \left(\frac{b'}{a}\right)^2\right)h_{R,\Vec{k}}'' + \left(2 \frac{a'}{a} + 2 \lambda \kappa^2 \frac{b'}{a} \frac{b''}{a}\right) h_{R,\Vec{k}}' + k^2 h_{R,\Vec{k}} \nonumber \\ &= \frac{2 k \epsilon \kappa^2}{a^2}  l_{\Vec{k}} \left( b'' h_{R,\Vec{k}}' + b' h_{R,\Vec{k}}'' + k^2 b' h_{R,\Vec{k}}\right)
\end{align}
where $l_{\Vec{k}} = +1$ and $l_{-\Vec{k}}=-1$. By~bringing all terms to one side, we get
\begin{equation}
    \begin{split}
        \left[1+\lambda\kappa^2 \left(\frac{b'}{a}\right)^2 + 2 \epsilon \kappa^2 l_{\Vec{k}} \frac{k}{a^2} b'\right] h_{L,\Vec{k}}'' &+ \left[2 \frac{a'}{a} + 2 \lambda \kappa^2 \frac{b'}{a} \frac{b''}{a} + 2 \epsilon \kappa^2 l_{\Vec{k}} \frac{k}{a^2} b''\right] h_{L,\Vec{k}}' \\
        &+ k^2 \left[1 + 2 \epsilon \kappa^2 l_{\Vec{k}} \frac{k}{a^2} b' \right] h_{L,\Vec{k}} = 0 
    \end{split}
\end{equation}
\begin{equation}
    \begin{split}
        \left[1+\lambda\kappa^2 \left(\frac{b'}{a}\right)^2 - 2 \epsilon \kappa^2 l_{\Vec{k}} \frac{k}{a^2} b'\right] h_{R,\Vec{k}}'' &+ \left[2 \frac{a'}{a} + 2 \lambda \kappa^2 \frac{b'}{a} \frac{b''}{a} - 2 \epsilon \kappa^2 l_{\Vec{k}} \frac{k}{a^2} b''\right] h_{R,\Vec{k}}' \\
        &+ k^2 \left[1 - 2 \epsilon \kappa^2 l_{\Vec{k}} \frac{k}{a^2} b' \right] h_{R,\Vec{k}} = 0
    \end{split} 
\end{equation}
These can be rewritten as
\begin{equation}
    h_{L,R,\Vec{k}}'' + \frac{1}{\kappa} P_{L,R,\Vec{k}} h_{L,R,\Vec{k}}' + k^2 Q_{L,R,\Vec{k}} h_{L,R,\Vec{k}} = 0
\end{equation}
where
\begin{align}
    P_{L,R,\Vec{k}} &= 2 \frac{\kappa \frac{a'}{a} + \lambda \kappa^3 \frac{b'}{a} \frac{b''}{a} + l_{L,R} \epsilon \kappa^3 l_{\Vec{k}} \frac{k}{a^2} b''}{1 + \lambda \kappa^2 \left(\frac{b'}{a}\right)^2 + 2 l_{\Vec{k}} \, l_{L,R} \epsilon\kappa^2 \frac{k}{a^2} b'} \\
    Q_{L,R,\Vec{k}} &= \frac{1 + 2 l_{\Vec{k}} \, l_{L,R} \epsilon \kappa^2 \frac{k}{a^2} b'}{1+\lambda\kappa^2 \left(\frac{b'}{a}\right)^2 + 2 l_{\Vec{k}} \, l_{L,R} \epsilon \kappa^2 \frac{k}{a^2} b'}
\end{align}\label{Qnumer}

The reader should note that the potentially dangerous modes with momenta, i.e.,~$k$, that lead to 
the cancellation of 
$$ 0 \simeq \lambda\kappa^2 \left(\frac{b'}{a}\right)^2 + 2 l_{\Vec{k}} \, l_{L,R}\epsilon \kappa^2 \frac{k}{a^2} b' $$
for $l_{\Vec{k}} l_{L,R}=-1$ in the denominators of $P_{L,\Vec{k}}, P_{R,\Vec{k}}$ are superhorizon (frozen) modes ($k < a H$), since they satisfy $k \simeq 0.053 \, a~H$; hence, this condition is never met by the subhorizon modes ($k > a H$) that contribute to the formation of the condensate~\cite{Dorlis:2024yqw}. In~any case, the~difference between $L,R$ modes (birefringence) persists for all subhorizon modes $k$. 

Taking into account that $k/a \lesssim \mu = M_s$, as~is appropriate for the UV cutoff of a string-effective theory like the current one (cf. \eqref{LMs}), we can estimate, using \eqref{Hbdot}, the following:
\begin{equation}
    \epsilon \kappa^2 \frac{k}{a} \dot{b} = \frac{\sqrt{2}}{12} \frac{a'}{\kappa} \kappa^2 \frac{k}{a} \dot{b} \lesssim \frac{10^{-1}}{M_S^2 M_{Pl}} \mu \dot{b} = \frac{10^{-1}}{M_S M_{Pl}} 10^{-1} M_{Pl} H = \frac{H}{M_S} 10^{-2} \lesssim 10^{-6}\label{c10}
\end{equation}
\begin{equation}\label{c11}
    \lambda \kappa^2 \dot{b}^2 = 10^{-1} a' \kappa^2 \dot{b}^2 = \frac{10^{-1}}{M_S^2 M_{Pl}^2} 10^{-2} M_{Pl}^2 H^2 = 10^{-3} \frac{H^2}{M_S^2} \lesssim 10^{-11}
\end{equation}
which allows us to approximate $Q$ as follows:
\begin{equation}\label{Qexp}
    \begin{split}
        Q & \simeq \left(1 + 2 l_{L,R} \epsilon \kappa^2 l_{\Vec{k}} \frac{k}{a^2} b'\right) \left(1-\lambda\kappa^2 \left(\frac{b'}{a}\right)^2 - 2 l_{\Vec{k}} \, l_{L,R} \epsilon \kappa^2 \frac{k}{a^2} b'\right) \\
        & \simeq 1 - \lambda\kappa^2 \left(\frac{b'}{a}\right)^2 
        \left(1 + 2 l_{\Vec{k}} \, l_{L,R} \epsilon \kappa^2 \frac{k}{a^2} b'\right)
        - 4 l_{\Vec{k}}^2 \, l_{L,R}^2 \epsilon^2 \kappa^4 \left(\frac{k}{a}\right)^2 \left(\frac{b'}{a}\right)^2 + \dots \,,
    \end{split}
\end{equation}

where $l_L = +1$ and $l_R = - 1$ and the $\dots$ denotes higher-order terms in our small-parameter expansion. Define
\begin{equation}
    z_{L,R} = e^{\frac{1}{2 \kappa} \int^\eta P_{L,R,\Vec{k}} d\tilde{\eta}}
\end{equation}
such that $\frac{1}{\kappa} P_{L,R,\Vec{k}} = 2 \frac{z^{\prime}_{L,R,\Vec{k}}}{z_{L,R,\Vec{k}}}$ and the equations become
\begin{equation}
    h_{L,R,\Vec{k}}'' + 2 \frac{z^{\prime}_{L,R,\Vec{k}}}{z_{L,R,\Vec{k}}} h_{L,R,\Vec{k}}' + k^2 Q_{L,R,\Vec{k}} h_{L,R,\Vec{k}} = 0
\end{equation}
Then, if~we define
\begin{equation}
    \psi_{L,R,\Vec{k}} = z_{L,R,\Vec{k}} h_{L,R,\Vec{k}},
\end{equation}
then, $\psi_{L,R,\Vec{k}}$ satisfies the equation
\begin{equation}
    \psi_{L,R,\Vec{k}}'' + \omega^2_{L,R,\Vec{k}} \psi_{L,R,\Vec{k}} = 0
    \label{psiosceqconf}
\end{equation}
where
\begin{equation}
    \omega^2_{L,R,\Vec{k}} = k^2 Q_{L,R,\Vec{k}} - \frac{z_{L,R,\Vec{k}}''}{z_{L,R,\Vec{k}}}
\end{equation}
We can perturbatively expand $Q_{L,R}$ and find that
\begin{equation}
    Q_{L,R,\Vec{k}} = 1 - \lambda \kappa^2 \left(\frac{b'}{a}\right)^2 + \mathcal{O}(\lambda\epsilon)
\end{equation}
or, in~cosmic time,
\begin{equation}
    Q_{L,R,\Vec{k}} = 1 - \lambda \kappa^2 \dot{b}^2 + \mathcal{O}(\lambda\epsilon)
\end{equation}
Furthermore, we have that
\begin{equation}
    \frac{z_{L,R,\Vec{k}}''}{z_{L,R,\Vec{k}}} = \frac{P_{L,R,\Vec{k}}^2}{4 \kappa^2} + \frac{P_{L,R,\Vec{k}}'}{2 \kappa}
\end{equation}
and we can show that
\begin{equation}
     \frac{z_{L,R,\Vec{k}}''}{z_{L,R,\Vec{k}}} = \frac{a''}{a} + \epsilon \kappa^2 k l_{\Vec{k}} \, l_{L,R} \Delta_1 + \lambda \kappa^2 b' (\Delta_1 + \Delta_2) + \mathcal{O}(\lambda\epsilon)
\end{equation}
where
\begin{equation}
    \Delta_1 = \frac{1}{a} \left[2 \left(\frac{a'}{a}\right)^2 \frac{b'}{a} - 2 \frac{a'}{a} \frac{b''}{a} - 2 \frac{b'}{a} \frac{a''}{a} + \frac{b'''}{a} \right]
\end{equation}
and
\begin{equation}
    \Delta_2 = \frac{1}{a}\left[- \left(\frac{a'}{a}\right)^2 \frac{b'}{a} + \frac{b'}{a} \frac{a''}{a} + \frac{a}{b'} \left(\frac{b''}{a}\right)^2\right]
\end{equation}
In cosmic time $t$, we have that
\begin{equation}
    \Delta_1 = a \left[\dddot{b} + H \ddot{b} - \left(H^2 + \frac{\ddot{a}}{a} \right)\dot{b}\right]
\end{equation}
and
\begin{equation}
    \Delta_2 = a \left[ \frac{\ddot{a}}{a} \dot{b} + H^2 \dot{b} + \frac{\ddot{b}^2}{\dot{b}} + 2H \ddot{b} \right]\,.
\end{equation}
Hence,

\begin{equation}
    \frac{z_{L,R,\Vec{k}}''}{z_{L,R,\Vec{k}}} = a^2 \Bigg[\left(H^2 + \frac{\ddot{a}}{a}\right) + \epsilon \kappa^2 \left(\frac{k}{a}\right) l_{\Vec{k}} \, l_{L,R} \left(\dddot{b} + H \ddot{b} - H^2 \dot{b} - \frac{\ddot{a}}{a} \dot{b} \right) + \lambda \kappa^2 \dot{b} \left(\dddot{b} + \frac{\ddot{b}^2}{\dot{b}} + 3 H \ddot{b}\right) \Bigg]
\end{equation}
Therefore, overall, we can evaluate $\omega^2_{L,R,\Vec{k}} = k^2 Q_{L,R,\Vec{k}} - \frac{z_{L,R,\Vec{k}}''}{z_{L,R,\Vec{k}}}$ in cosmic time:

\begin{equation}
    \begin{split}
        \omega_{L,R,\Vec{k}}^2 = a^2 \Bigg[\left(\frac{k}{a}\right)^2 \left(1 - \lambda \kappa^2 \dot{b}^2 \right) - \left(H^2 + \frac{\ddot{a}}{a}\right) &- \epsilon \kappa^2 \left(\frac{k}{a}\right) l_{\Vec{k}} \, l_{L,R} \left(\dddot{b} + H \ddot{b} - H^2 \dot{b} - \frac{\ddot{a}}{a} \dot{b} \right) \\
        &- \lambda \kappa^2 \dot{b} \left(\dddot{b} + \frac{\ddot{b}^2}{\dot{b}} + 3 H \ddot{b}\right) \Bigg]
    \end{split}
\end{equation}
This can be rewritten as
\begin{equation}
    \begin{split}
        \omega_{L,R,\Vec{k}}^2 = a^2 \Bigg[\left(\frac{k}{a}\right)^2 - \left(H^2 + \frac{\ddot{a}}{a}\right) &- \epsilon \kappa^2 \left(\frac{k}{a}\right) l_{\Vec{k}} l_{L,R} \left(\dddot{b} + H \ddot{b} - H^2 \dot{b} - \frac{\ddot{a}}{a} \dot{b} \right) \\
        &- \lambda \kappa^2 \dot{b} \left(\dddot{b} + \frac{\ddot{b}^2}{\dot{b}} + 3 H \ddot{b} + \left(\frac{k}{a}\right)^2 \dot{b} \right) \Bigg]
    \end{split}
\end{equation}
and so we have that (\ref{psiosceqconf}) can be written as
\begin{equation}
    \psi_{L,R,\Vec{k}}'' + \omega^2_{L,R,\Vec{k}} \psi_{L,R,\Vec{k}} = 0 \Rightarrow \ddot{\psi}_{L,R,\Vec{k}} + H \dot{\psi}_{L,R,\Vec{k}} + \frac{\omega^2_{L,R,\Vec{k}}}{a^2} \psi_{L,R,\Vec{k}} = 0
\end{equation}
i.e.,
\begin{align}
    \ddot{\psi}_{L,R,\Vec{k}} + H \dot{\psi}_{L,R,\Vec{k}} + \Bigg[\left(\frac{k}{a}\right)^2 - \left(H^2 + \frac{\ddot{a}}{a}\right) &- \epsilon \kappa^2 \left(\frac{k}{a}\right) l_{\Vec{k}} \, l_{L,R} \left(\dddot{b} + H \ddot{b} - H^2 \dot{b} - \frac{\ddot{a}}{a} \dot{b} \right) \nonumber \\
    &- \lambda \kappa^2 \dot{b} \left(\dddot{b} + \frac{\ddot{b}^2}{\dot{b}} + 3 H \ddot{b} + \left(\frac{k}{a}\right)^2 \dot{b} \right) \Bigg]\, \psi_{L,R,\Vec{k}} = 0\,.
\end{align}
For $\dot{b} \simeq {\rm const}$ and $\frac{\ddot{a}}{a} \simeq H^2$, we have that%

{\begin{equation}
    \ddot{\psi}_{L,R,\Vec{k}} + H \dot{\psi}_{L,R,\Vec{k}} + \Bigg[\left(\frac{k}{a}\right)^2 - 2H^2 + 2 \epsilon \kappa^2 \left(\frac{k}{a}\right) l_{\Vec{k}} \, l_{L,R} H^2 \dot{b} - \lambda \kappa^2 \left(\frac{k}{a}\right)^2 \dot{b}^2 \Bigg]\, \psi_{L,R,\Vec{k}} = 0\,.
\end{equation}}
or, equivalently
{\begin{equation}
    \ddot{\psi}_{L,R,\Vec{k}} + H \dot{\psi}_{L,R,\Vec{k}} + \Bigg[\left(\frac{k}{a}\right)^2 \left(1 - \lambda \kappa^2 \dot{b}^2 \right) - 2H^2 + 2 \epsilon \kappa^2 \left(\frac{k}{a}\right) l_{\Vec{k}} \, l_{L,R} H^2 \dot{b} \Bigg]\, \psi_{L,R,\Vec{k}} = 0\,.
\end{equation}}
We rescale $\frac{1}{\tilde{a}} = \frac{\sqrt{1-\lambda\kappa^2 \dot{b}^2}}{a}$ (cf. \eqref{arescale}) and get
{\begin{equation}
    \ddot{\psi}_{L,R,\Vec{k}} + H \dot{\psi}_{L,R,\Vec{k}} + \Bigg[\left(\frac{k}{\tilde{a}}\right)^2 - 2H^2 + 2 \epsilon \kappa^2 \frac{1}{\sqrt{1-\lambda\kappa^2 \dot{b}^2}}\left(\frac{k}{\tilde{a}}\right) l_{\Vec{k}} \, l_{L,R} H^2 \dot{b} \Bigg]\, \psi_{L,R,\Vec{k}} = 0\,,
\end{equation}}
which, upon~expanding, becomes
{\begin{equation}
    \ddot{\psi}_{L,R,\Vec{k}} + H \dot{\psi}_{L,R,\Vec{k}} + \Bigg[\left(\frac{k}{\tilde{a}}\right)^2 - 2H^2 + 2 \epsilon \kappa^2\left(\frac{k}{\tilde{a}}\right) l_{\Vec{k}} \, l_{L,R} H^2 \dot{b} + \mathcal{O}(\epsilon \lambda) \Bigg]\, \psi_{L,R,\Vec{k}} = 0\,.
\end{equation}}%
This is nothing other than the corresponding wave equations of chiral GW linear polarizations of~\cite{Dorlis:2024yqw}, expressed in terms of the rescaled scale factor $a^\lambda$ \eqref{arescale}. Following the canonical quantization approach of that work, then, in~the evaluation of the gravitational CS condensate for a weak $\dot b$, leads to \eqref{RCSconformal2} in the~text.

\bibliographystyle{apsrev4-2}
\bibliography{MP2025refs}  

\end{document}